\documentclass[reprint,a4paper]{article}
\usepackage[affil-it]{authblk}

\usepackage{setspace}
\usepackage{color}
\usepackage[dvipsnames]{xcolor}
\usepackage{amsmath}
\usepackage{amsfonts}
\usepackage{subcaption}
\usepackage{comment}
\usepackage{booktabs}
\usepackage{verbatim}
\usepackage{amssymb}
\usepackage{graphicx}
\usepackage{hyperref}
\usepackage[nameinlink]{cleveref}
\usepackage{cite}
\usepackage{bbold}
\usepackage{dsfont}
\usepackage{url}
\usepackage{caption}
\usepackage{subcaption}
\usepackage[top=2cm, bottom=2cm, left=3cm, right=3cm]{geometry}
\usepackage{upgreek}

\providecommand{\keywords}[1]
{
  \small
  \textbf{\textit{Keywords---}} #1
}

\definecolor{darkgreen}{rgb}{0,0.35,0}

\newcommand{\CECs}{Centro de Estudios Cient\'{\i}ficos (CECs), Casilla 1469, Valdivia, Chile}
\newcommand{\USS}{Universidad San Sebasti\'{a}n, sede Valdivia, General Lagos 1163, Valdivia 5110693, Chile}
\newcommand{\UNAPa}{Facultad de Ciencias, Universidad Arturo Prat, Avenida Arturo Prat Chacón 2120, 1110939, Iquique, Chile}
\newcommand{\UNAPb}{Instituto de Ciencias Exactas y Naturales, Universidad Arturo Prat, Playa Brava 3256, 1111346, Iquique, Chile}
\newcommand{\UACh}{Instituto de Ciencias F\'isicas y Matem\'aticas, Universidad Austral  de Chile, Casilla 567, Valdivia, Chile}
\newcommand{\UCharles}{IPNP - Faculty of Mathematics and Physics, Charles University, V Hole\v{s}ovi\v{c}k\'ach 2, 18000 Prague 8, Czech Republic}

\begin{document}

\title{Exact mapping from the $(3+1)$-dimensional Skyrme model to the $(1+1)$-dimensional sine-Gordon theory and some applications}

\author[1,2]{Fabrizio Canfora\thanks{fabrizio.canfora@uss.cl}}
\author[3,4]{Marcela Lagos\thanks{marclagos@unap.cl}}
\author[5,6]{Pablo Pais\thanks{pais@ipnp.troja.mff.cuni.cz}}
\author[6]{Aldo Vera\thanks{aldo.vera@uach.cl}}

\affil[1]{\USS}
\affil[2]{\CECs}
\affil[3]{\UNAPa}
\affil[4]{\UNAPb}
\affil[5]{\UACh}
\affil[6]{\UCharles}

\date{}

\maketitle

\begin{abstract}
A remarkable exact mapping, valid for low-enough energy scales and close to a sharp boundary distribution of hadronic matter, from the $(3+1)$-dimensional Skyrme model to the sine-Gordon theory in $(1+1)$ dimensions in the attractive regime is explicitly constructed. Besides the intrinsic theoretical interest to be able to describe the prototype of nonintegrable theories (namely, quantum chromodynamics in the infrared regime) in terms of the prototype of integrable relativistic field theories (namely, sine-Gordon theory in $(1+1)$ dimensions), we will show that this mapping can be extremely useful to analyze both equilibrium and out-of-equilibrium features of baryonic distributions in a cavity.
\end{abstract}

\keywords{Skyrme model; Nuclear Physics; Non-equilibrium QFT}F

\newpage

\section{Introduction}

\label{Intro} 

In condensed matter physics, the analysis both of the phase diagram as well as of out-of-equilibrium features (such as the entanglement and its evolution in many-body physics: see \cite{entanglement1a,entanglement2,entanglement3,entanglement5c,entanglement5d,entanglement6,entanglement7,entanglement8,entanglement9aa1,entanglement9a1,entanglement9a2,entanglement9b,entanglement9c,entanglement9d,entanglement10}%
, detailed reviews are \cite{entanglement11,entanglement12,entanglement13,entanglement13a,entanglement13b,entanglement14,entanglement15aa1,entanglement15a}%
) is one of the most important topics, both theoretically and experimentally. These topics are at the crossroad between statistical physics, quantum computation and quantum field theory. In low-dimensional systems, many powerful exact results have been derived. For instance, the references mentioned above indicate that the asymptotic entanglement of a large subsystem has been related to thermodynamic entropy in a stationary
state. It has also been possible to connect the growth of entanglement with the capability of a classical computer to simulate nonequilibrium quantum systems with matrix product states. Integrable models in $(1+1)$ dimensions offer a unique window on interacting systems, allowing a detailed understanding of the time evolution of entanglement \cite{entanglement9aa1,entanglement9a1,entanglement9a2,entanglement9b,entanglement9c,entanglement9d}.

These concepts are essential in relativistic quantum field theory (QFT) as well, with implications from black hole physics to scattering amplitudes and quantum chromodynamics (QCD) \cite{entanglementQFT1,entanglementQFT2,entanglementQFT3,entanglementQFT4}. However, except for conformal field theory and integrable models \cite{SG,SG0,SG1b,SG2,SG3c,SG4b,SG5,SG6,Koch:2023xuw,SG7,SGfinite1,SGfinite2,SGKT1a,SGKT1b,SGKT3,SGKT4}, at first glance, it looks impossible to obtain exact results either on the phase diagram (such as the ones in \cite{SG,SG0,SG6,Koch:2023xuw}) or on entanglement dynamics (such as the ones in \cite{entanglement9aa1,entanglement9a1,entanglement9a2}) of strongly interacting QFT. A theoretical dream is to find a mapping that allows to use these powerful results available in integrable field theories in the analysis of both equilibrium and out-of-equilibrium features of the low energy limit of QCD, where perturbation theory is useless.

In the present work, we will show that, at least in some cases, such a dream can be achieved. In particular, we will construct a mapping that enables ne to derive results similar to \cite{entanglement9aa1,entanglement9a1,entanglement9a2,SG,SG0,SG6,Koch:2023xuw} in the case of a distribution of Baryonic matter close to its boundary. Since only refined numerical techniques are commonly employed in analyzing the phase diagram of QCD at low temperatures and finite baryon densities \cite{newd3a,newd4,newd5,newd6}, the present analytic results are highly relevant and produce novel robust predictions which can, in principle, be tested.

The starting point is the Skyrme theory, which (at leading order in the 't Hooft expansion \cite{Gerard1,largeN1,largeN2}) represents the low-energy limit of QCD. The dynamical field of the Skyrme action \cite{skyrme1} is a $SU(N)$-valued scalar field $U$ (here, we will consider the two-flavors case). This action possesses both small excitations describing pions and topological solitons describing baryons \cite{Lizzi,witten0a,ANW,Guadagnini,aprox4}, being the baryonic charge a topological invariant. Skyrme theory has always been considered the prototype of nonintegrable models where the powerful nonperturbative results available in quantum many-body physics \cite{Eisert:2014jea,Kormos_2016,Meldgin_2016,QMBPRX1,QMBPRX2} cannot be applied.
However, the techniques developed in \cite{56a0,56b1,56c,crystal1a,crystal1b,crystal3} allowed for the first time a successful analytic description of nonhomogeneous baryonic condensates at finite baryon density, in good qualitative agreement with the available phenomenological results in \cite{pasta10} and references therein. This framework allows describing $(3+1)$-dimensional baryonic layers confined in a cavity in terms of the sine-Gordon theory (SGT), which is the prototype of a relativistic integrable field theory in $(1+1)$ dimensions. Hence, the results in \cite{SG1b,SG2,SG3c,SG4b,SG5,SG6,Koch:2023xuw,SG7} developed for the SGT at finite temperatures and in \cite{entanglement9aa1,entanglement9a1,entanglement9a2,entanglement9b,entanglement9c,entanglement9d} for SGT out-of-equilibrium produce novel analytic results in the low-energy limit of QCD at low temperatures.


\section{Summary of the results}


The action for the $SU(2)$-Skyrme model is given by
\begin{align}  \label{actionskyrme1}
I[U]\ &= \ \frac{K}{4} \, \int_{\mathcal{M}} d^{4}x \, \sqrt{-g}\ \text{Tr} \left( R_{\mu }R^{\mu }+\frac{\lambda }{8}G_{\mu \nu }G^{\mu \nu }\right) \;,
\\
R_{\mu } &= U^{-1}\nabla _{\mu }U=R_{\mu }^{a}t_{a}\ ,\quad G_{\mu\nu}=[R_{\mu },R_{\nu }]\;,  \notag
\end{align}
where $U(x)\in SU(2)$, $g$ is the metric determinant, $\nabla _{\mu }$ is the partial derivative, and $t_{a}=i\sigma _{a}$ are the generators of the $SU(2)$ Lie group, being $\sigma _{a}$ the Pauli matrices. The space-time manifold is split in $\mathcal{M}=\mathds{R}\times\Sigma$, and the Skyrme couplings $K$ and $\lambda$ are positive constants fixed experimentally.

The topological current $J^{\mu}$ and the baryonic charge $Q_{B}$ are defined, respectively, as
\begin{align}
J^{\mu}\ & =\ \epsilon ^{\mu \nu \alpha \beta}\,\text{Tr}(R_{\nu }R_{\alpha}R_{\beta })\;,  \notag  \label{currents} \\
Q_{B}\ & =\ \frac{1}{24\pi ^{2}}\int_{\Sigma }J^{0}\;.
\end{align}
where the last integral is performed on $\Sigma$. Geometrically, a nonvanishing $J^{\mu }$ measures the ``genuine three-dimensional nature'' of the configuration since (in order to have $J^{\mu }\neq 0$, at least locally) $U$ must encode three independent degrees of freedom. For instance, if two of the three degrees of freedom needed to describe $U$ depend on the same coordinate, then $J^{\mu}$ vanishes identically.

We want to analyze the intriguing phenomena that occur when a finite amount of baryons live within a cavity of finite spatial volume $V=8\pi^{2}L^{2}L_{x}$; therefore, we consider the metric of a box
\begin{equation}
ds^{2}=-dt^{2}+dx^{2}+L^{2}(d \mathfrak{y} ^{2}+d \mathfrak{z} ^{2}) \;,
\label{Box}
\end{equation}
where $L_{x}$ and $L$ are constants representing the size of the box in the directions longitudinal and orthogonal to $x$, respectively. The coordinates have the following ranges (see \cite{56a0,56b1}):
\begin{equation}  \label{ranges}
0\leq x\leq L_{x} \, , \quad \ 0\leq \mathfrak{y} \leq 2\pi \, , \quad \
0\leq \mathfrak{z} \leq 4\pi \;.
\end{equation}
Note that the coordinate $x$ has a length dimension, while the other two coordinates, $\mathfrak{y}$ and $\mathfrak{z}$, are dimensionless (since the length scale $L$ has been explicitly shown in the metric). This helps to analyze the interplay between the two scales, $L$ and $L_{x}$, in the
following computations. In order to apply the known results on SGT mentioned above, the limit $L_{x}\gg L$ has to be considered. In this case, one would describe a sort of ``hadronic wire'' (a cavity much longer in one spatial direction than in the other two). We will choose $\mathfrak{y}$ and $\mathfrak{z}$ as the ``homogeneous coordinates''. When $L_{x}$ is not large, one has to use the exact available results on SGT on a finite interval \cite{SGfinite1,SGfinite2}.

The Skyrme action and the corresponding field equations can be written explicitly in terms of the $SU(2)$-valued field (as any element of $SU(2)$ can be written in the Euler representation)
\begin{equation}
U=\exp \left( t_{3}\,F\right) \exp \left( t_{2}\,H\right) \exp \left(
t_{3}\,G\right) \ ,  \label{I2}
\end{equation}%
where $F=F\left( x^{\mu }\right) $, $G=G\left( x^{\mu }\right) $ and $H=H\left( x^{\mu }\right) $ are the three scalar degrees of freedom of the Skyrme field (traditionally, in this parametrization, the field $H$ is called \textit{profile}).

It is well known that many baryonic distributions, especially at low energies, possess a sharp boundary. Namely, the energy and baryon densities decay exponentially fast to zero. Thus, in practice, one can define a surface that separates the region where the energy and baryon densities are different from zero, from the region where these densities vanish. \textit{In a neighborhood of any point close to such a surface}, the energy and baryon density can only depend on the spatial coordinate orthogonal to the surface and on time (we will comment more on this point in the following sections). This is why it is so interesting to study hadronic distributions that are homogeneous in two spatial directions. The analysis of the energy-momentum tensor shows that the only Ansatz able to describe energy and baryon densities homogeneous in two spatial directions \cite{56c} is \begin{equation}
H(x^{\mu })=H(t,x)\ ,\quad F(x^{\mu })=\frac{q}{2}\,\mathfrak{y}\ ,\quad
G(x^{\mu })=\frac{p}{2}\,\mathfrak{z}\ .  \label{Ansatz01}
\end{equation}%
The above Ansatz has several remarkable properties. First, the three coupled nonlinear Skyrme field equations reduce consistently to just one partial differential equation (PDE) for the profile; such a PDE is the sine-Gordon equation in $(1+1)$ dimensions. Second, the topological charge density is nontrivial, leading to an arbitrarily high baryon number. Third, suppose the energy/temperature scale is less than $1/L$. In that case, the only relevant degrees of freedom are fluctuations $\delta H(t,x)$ of $H(t,x)$, which only depend on $t$ and $x$, as all the other possible fluctuations of the field $U $ have energies larger than $1/L$. These features of hadronic layers will be the key to deriving novel properties and making robust predictions on their behavior at finite density using known results in the literature on SGT.

\subsection{Effective sine-Gordon theory}

We will consider $p=q$, with $B\equiv p^{2}>0$, to reduce the complexity of the formulas. Nevertheless, all the present results generalize easily to cases where $p$ and $q$ are arbitrary integers. The on-shell Lagrangian and the baryon density, $\rho_{B}\,\equiv\,J_{0}$, corresponding to the Ansatz in \Cref{Ansatz01}, give rise to the following effective SGT (we dropped out a constant term in the action that does not affect the theory)
\begin{equation}
I_{\text{SG}}\ =\ \int L_{\text{SG}}\,dtdx\, = \,\int \left(-\frac{1}{2} \partial ^{\mu}\varphi \partial _{\mu }\varphi + M_{0}\left(\cos(\beta\,\varphi)-1\right) \right)\, dt dx\ , \qquad \varphi \ = \frac{4}{\beta} H \;,  \label{id1}
\end{equation}
with
\begin{gather}
\quad M_{0}=\frac{\pi ^{2}K}{8L^{2}}B^{2}\lambda \;, \quad \beta \ =\ \frac{2}{\pi \left[ K\left( 2L^{2}+B\,\lambda \right) \right] ^{1/2}} \;,
\label{id}
\end{gather}
where constant terms have been discarded. From the above, the complete set of Skyrme field equations are reduced to the sine-Gordon equation for the $\varphi$ field,
\begin{equation}  \label{EqH}
\Box \varphi - M_0 \beta \sin(\beta \varphi) \ = 0 \ , \qquad \Box \equiv -\partial_t^2 + \partial_x^2 \;,
\end{equation}
which, in the static case, can be reduced to a quadrature
\begin{equation}  \label{EqH0}
dx=\frac{d\varphi}{\sigma(E_0,\varphi)} \ , \qquad \sigma(E_0,\varphi)=\pm \biggl( \frac{E_0}{L^2} -2M_0 \cos(\beta \varphi) \biggl)^{\frac{1}{2}} \;,
\end{equation}
where $E_0>2L^{2}M_{0}$ is an integration constant.

Hence, one can deduce many intriguing analytic results on this $(3+1)$-dimensional distribution of baryonic matter confined in a cavity by using the classic SGT results in \cite{entanglement9aa1,entanglement9a1,entanglement9a2,entanglement9b,entanglement9c,entanglement9d}, provided the energy/temperature scale is less than $1/L$. The boundary conditions for $H$ (or $\varphi $) are fixed by requiring that the baryonic charge in \Cref{currents} is the integer $B$,
\begin{equation}
\begin{array}{l}
H(t,0)=n\frac{\pi }{2}\, \\
H(t,L_{x})=(m+\frac{n+1}{2})\,\pi
\end{array}%
\ \ \ \ \ \Leftrightarrow \ \ \ \ \
\begin{array}{l}
\varphi (t,0)=2n\frac{\pi }{\beta }\, \\
\varphi (t,L_{x})=\frac{2}{\beta }\,\left( n+(2m+1)\right) \,\pi
\end{array}
\;,  \label{BC1}
\end{equation}
where $n$ and $m$ are integers.

If $H$ satisfies either $H(t,L_{x})=H(t,0) + n\,\pi$, or $H(t,L_{x}) + H(t,0)= m\,\pi$, for $n,m\in\mathds{Z}, \quad \forall\,t$, the total baryonic charge vanishes. Despite this, such configurations are interesting anyway, since the baryonic density is nontrivial and one can have bound states of two layers.

\begin{figure}[tbp]
\begin{center}
\includegraphics[width=0.5\textwidth,angle=0]{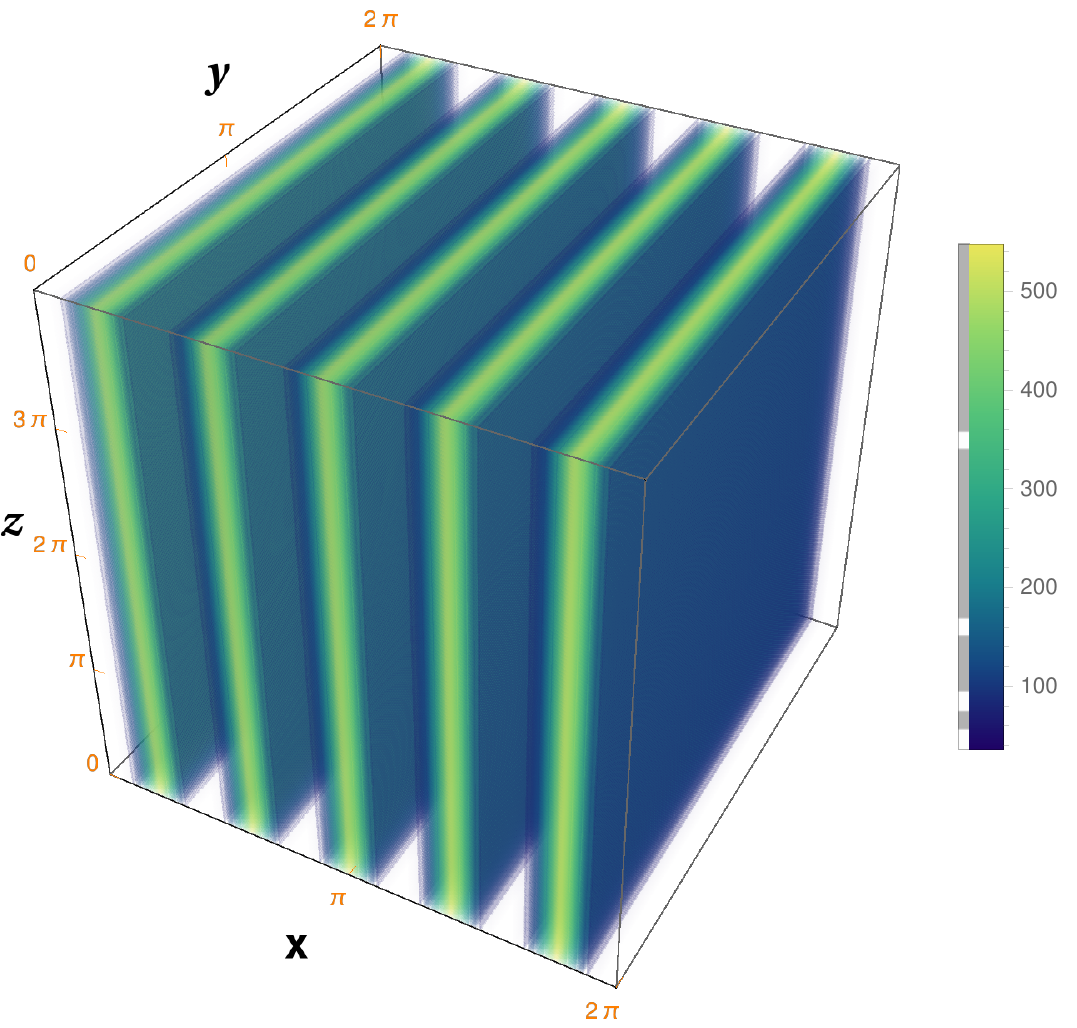}
\end{center}
\caption{Energy density of baryonic layers in a cavity.}
\label{Figure_ED}
\end{figure}

\Cref{Figure_ED} shows the energy density of a configuration with baryonic charge $B=16$ describing baryonic layers in a cavity, where we have considered $n=0$ and $m=2$ for the boundary conditions in \Cref{BC1}.
For static configurations, $\varphi (t,x)=\varphi (x)$, the integration constant $E_{0}$, according to \Cref{ranges,id,EqH0,BC1}, is fixed by \begin{equation}  \label{BC2}
\pm \int_{0}^{2 \pi/\beta}\frac{d\varphi }{ [ E_0-2 L^2 \, M_0 \, \cos(\beta \, \varphi) ]^{\frac{1}{2}}}=\frac{L_{x}}{L}\ \;,
\end{equation}
where we have taken, for simplicity, $n=m=0$ in \Cref{BC1}. It is easy to see that the above equation always has a solution if $L_{x}$ is finite. Indeed, if $L_{x}$ is small (compared to $L$) one can take a large $E_{0}$ to make the left-hand side of \Cref{BC2} small as well. If $L_{x}$ is large (but not divergent), one can have the left-hand side of \Cref{BC2} large by choosing
\begin{equation}
E_{0} = 2L^{2}\,M_{0} + \varepsilon \ , \qquad 0<\varepsilon \ll 1 \ ,
\end{equation}%
so that the denominator comes close to have a zero when $\varphi=0$, or $\varphi = 2\pi\,\beta$. The $\frac{L_{x}}{L}\rightarrow \infty$ case corresponds to the limit in which $\varepsilon =0$.

The results in \cite{entanglement9aa1,entanglement9a1,entanglement9a2,entanglement9b,entanglement9c,entanglement9d, entanglement15aa1,entanglement15a,SG1b,SG2,SG3c,SG4b,SG5,SG6,Koch:2023xuw,SG7,SGfinite1,SGfinite2} can now be used. The presence of both the two directions $\mathfrak{y}$ and $\mathfrak{z}$ orthogonal to $x$ and the baryonic number manifests itself in the effective sine-Gordon couplings $M_{0}$ and $\beta $. Thus, according to \Cref{id}, whether the effective SGT describing the hadronic distribution is in the attractive phase ($\mathbf{\mathit{AP}}$), repulsive phase ($\mathbf{\mathit{RP}}$) or critical free-Fermion phase ($\mathbf{\mathit{FFP}}$), can be deduced from the factor $\frac{\beta ^{2}}{4\pi}$, explicitly
\begin{equation}
\mathbf{\mathit{AP}}:\frac{1}{\pi ^{3}K\left(2L^{2}+ B\, \lambda \right) }<1 \ , \ \ \mathbf{\mathit{RP}}:\frac{1}{\pi ^{3}K\left( 2L^{2}+ B\,\lambda \right) }>1 \ , \ \ \mathbf{\mathit{FFP}}:\frac{1}{\pi ^{3}K\left( 2L^{2} + B\,\lambda \right) }=1 \;.
\end{equation}
Since $B$ is a positive integer, $K\lambda $ is around $1/6$ (see Ref. \cite{ANW}) and $\pi ^{3}>27$, then the effective theory is always in the attractive regime and, therefore, the results in Refs. \cite{entanglement9a1,entanglement9a2} can be applied here. In practice, the number of breathers
\begin{equation}
\frac{1}{\xi }=\frac{1-\beta ^{2}}{\beta ^{2}} \;,  \label{id0}
\end{equation}
which is either $\frac{1}{\xi }-1$ or $\left[ \frac{1}{\xi } \right]$ depending on whether $\frac{1}{\xi }\in\mathbb{Z}$ or not, is always bigger than two, already for $B \geq 4$.

\subsection{In and out-of-equilibrium implications}

The description of $(3+1)$-dimensional hadronic layers in a cavity presented above in terms of the SGT in $(1+1)$ dimensions offers unprecedented possibilities. When the energy/temperature scale is low enough, the equilibrium and out-of-equilibrium properties of these configurations can be computed using the effective SGT with coupling constants in \Cref{id1,id}. It is worth reminding that, if one is close enough to the boundary of a baryonic distribution of matter (such that, in one spatial direction, the energy and baryon densities drop very rapidly to a very small value, while in the two orthogonal directions, the energy and baryon densities are almost homogeneous), then the present description in terms of SGT is actually generic.

\textit{The first type} of exact results, which can be ``imported" from SGT in the analysis of baryonic distributions in a cavity, it has to do with the mass spectrum of the theory, with the equilibrium correlation functions at finite temperatures (but smaller than $1/L$) \cite{SG,SG0} and with the phase diagram \cite{entanglement5c,entanglement5d,entanglement13,entanglement13a,entanglement13b,entanglement15aa1,entanglement15a}. If $L_{x}/L$ is large enough, all such analytic results can be applied directly to the effective SGT in \Cref{id1,id}. In this way, one can get the exact excitations' spectrum of the hadronic distribution as well as the corresponding low temperatures' correlation functions.

\textit{The second type} is the computation of the entanglement entropy \cite{entanglement9b,SG,SG0}. In particular, the above references together with the present mapping imply that the entanglement entropy $S_{E}$ of hadronic layers confined in a cavity is $S_{E}=\frac{c_{\text{eff}}}{3}\ln \left(\frac{l}{a}\right)$; where $a$ is the UV cutoff beyond which the Skyrme model is not valid anymore, and $l$ is the size of the finite interval in the $x$ direction, of which we are computing the corresponding entanglement entropy. The effective central charge $c_{\text{eff}}$ of the theory can be estimated following Refs. \cite{SG,SG0}.

\textit{The third} (and, perhaps, most surprising) \textit{type} has to do with out-of-equilibrium properties, such as the dynamics of entanglement entropy after a quantum quench and the Loschmidt echo. These quantities are entirely out of reach of any standard perturbative approach based on QCD in $(3+1)$ dimensions, especially at finite baryon density. Nevertheless, the present mapping allows using directly the results in Refs. \cite{entanglement9aa1,entanglement9a1,entanglement9a2,entanglement9b,entanglement9c,entanglement9d,entanglement14,entanglement15aa1,entanglement15a}. These together with the present mapping, allow concluding that one can obtain exact analytical predictions for the time evolution of the entanglement in generic hadronic layers confined in a three-dimensional cavity. Even more, following Ref. \cite{entanglement9aa1} (as in the present case, the effective SGT is always in the attractive regime with more than two breathers), the von Neumann and Renyi entropies display undamped oscillations in time, whose frequencies can be taken exactly from Refs. \cite{entanglement9a1,entanglement9a2}. For instance, the quantum quench can be realized by either changing $B$ or $L$ (moreover, the number of oscillatory modes grows with $B$). Finally, following Ref. \cite{entanglement15aa1}, the Loschmidt amplitude, the fidelity and work distribution can be computed explicitly.


\section{Technical details}


\subsection{Euler angles parametrization and energy-density}

\label{section_technical_Euler}

The field equations of the system, obtained varying the action in \Cref{actionskyrme1} with respect to the $U$ field, are
\begin{equation}  \label{Eqoskyrme}
\nabla _{\mu }\left( R^{\mu }+\frac{\lambda }{4}[R_{\nu },G^{\mu \nu}]\right) =0 \;.
\end{equation}
Equations (\ref{Eqoskyrme}) are generally a set of three coupled nonlinear partial differential equations. The energy-momentum tensor of the theory is given by
\begin{equation}  \label{TMUNU}
T_{\mu \nu }=-\frac{K}{2}\text{Tr}\left[ R_{\mu }R_{\nu }-\frac{1}{2} g_{\mu\nu }R^{\alpha }R_{\alpha }+\frac{\lambda }{4}\left( g^{\alpha \beta
}G_{\mu\alpha }G_{\nu \beta }-\frac{1}{4}g_{\mu \nu }G_{\sigma \rho} G^{\sigma \rho}\right) \right] \;.
\end{equation}

We are interested in finite density effects: in particular, we want to describe hadronic layers confined to a cavity, where the Euler angles parametrization for the Skyrme field is particularly convenient. The wording ``hadronic layers'' refers to distributions of energy density $T_{00}$ and baryon density $J^{0}$, which are homogeneous in two spatial coordinates. Still, they depend nontrivially on the third spatial coordinate and on time. The interest in such configurations lies, at the very least, in the following facts.

\begin{figure}[tbp]
\begin{center}
\includegraphics[width=0.80\textwidth,angle=0]{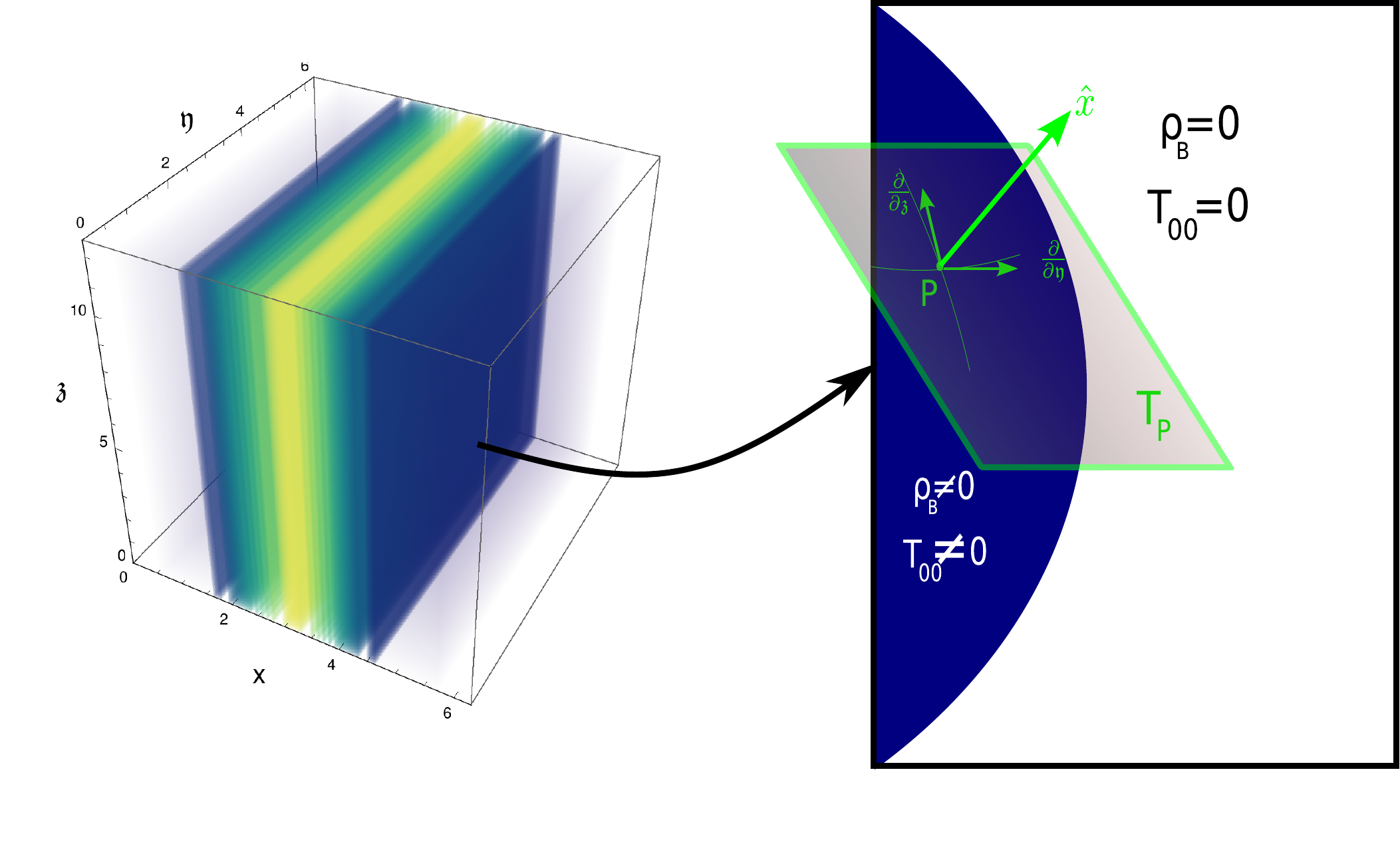}
\end{center}
\caption{A schematic representation of the baryonic layer in the coordinates $(x,\mathfrak{y},\mathfrak{z})$ of (\protect\ref{ranges}). Taking a specific solution, if we make a zoom-in, we find a surface perpendicular to the direction where $\protect\rho_{B}$ drops very rapidly to zero, in this case, $\hat{x}$, separating two distinct regions: $\protect\rho_{B}\neq0$ (interior) and $\protect\rho_{B}=0$ (exterior). As the profile $H(x,t)$ does not depend on the spatial dimensions $\mathfrak{y}$ and $\mathfrak{z}$, at point $p$, the tangent space $T_{p}$ (where the coordinate basis $\{\frac{\partial}{\partial_{\mathfrak{y}}},\frac{\partial}{\partial_{\mathfrak{z}}}\}$ belongs to) is \emph{approximately} the baryonic layer. }
\label{Figure_Approximation}
\end{figure}

\textit{First}, it is possible to define the ``boundary'' of the distribution of nuclear matter as the surface in space where the energy density and baryon density drop exponentially fast to zero (or to a very small constant value). In a small enough neighborhood of a point on such boundary, the energy density and baryon density will not depend on the two spatial coordinates tangent to the boundary. In contrast, they will depend very sensitively on the spatial coordinate orthogonal to the boundary (since both $T_{00}$ and $J^{0}$ drop rapidly to zero along this spatial direction). Hence, the configurations discussed here are quite generic and useful close \textit{to a sharp boundary distribution of hadronic matter} (see Figure \ref{Figure_Approximation}).

\textit{Second}, such structures are known to appear in numerical simulations at finite baryon density, and there is robust phenomenological evidence supporting their presence in neutron stars (see \cite{pasta10} and references therein). Consequently, the starting point is the metric of a cavity in \Cref{Box} and \Cref{ranges}.

\begin{figure}[tbp]
\begin{center}
\includegraphics[width=0.35\textwidth,angle=0]{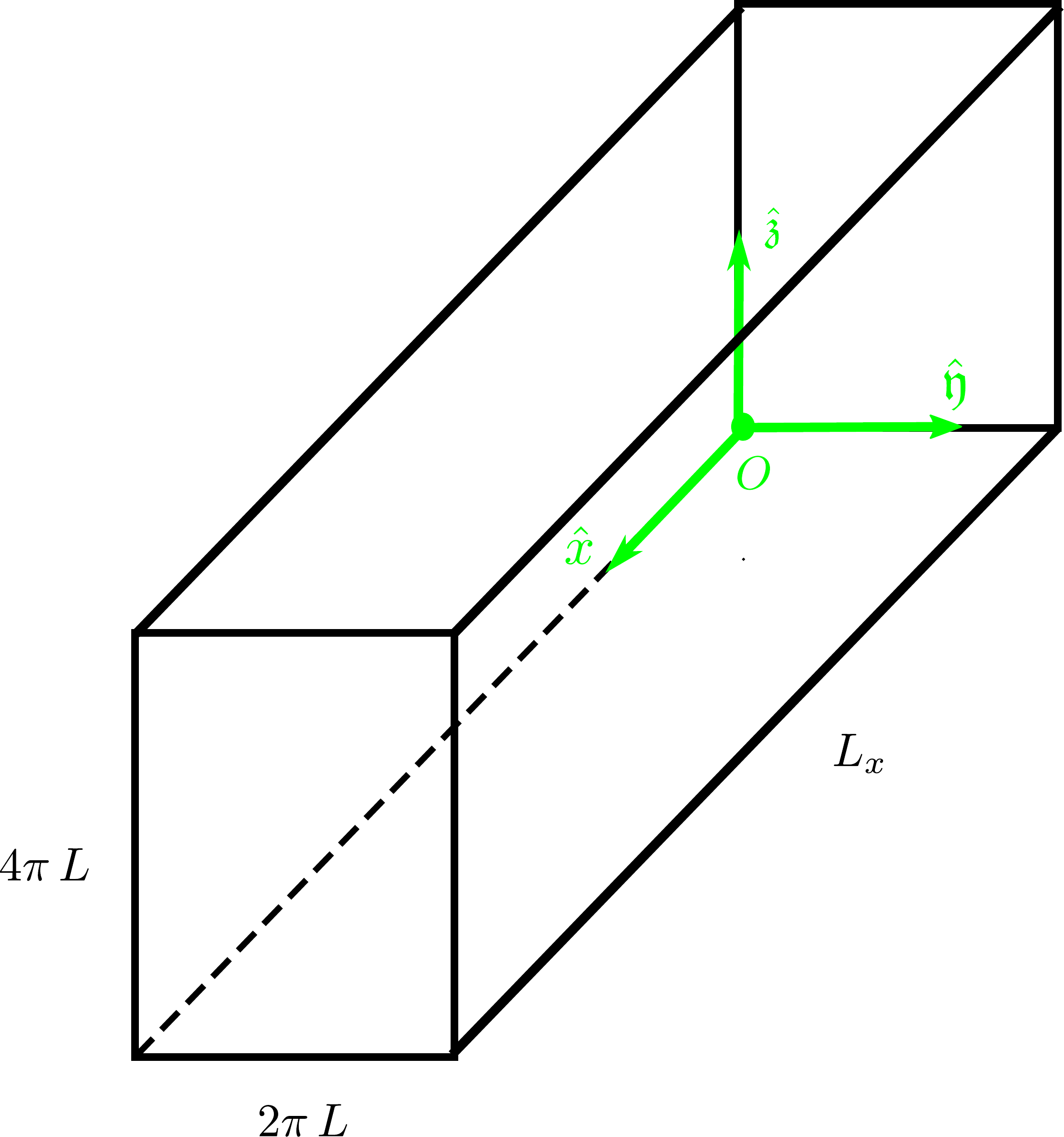}
\end{center}
\caption{The range of the coordinates $(x,\mathfrak{y},\mathfrak{z})$ of (\protect\ref{ranges}) represents a \textquotedblleft hadronic wire\textquotedblright . The longer edge has a length of $L_{x}$, while the other two have a size of $2\protect\pi \,L$ and $4\protect\pi \,L$. The
origin $O$ is on one of its corners, and the coordinates axes are represented in light green.}
\label{Figure_Cylinder}
\end{figure}

In the present setting, to apply the powerful results on the sine-Gordon theory (SGT) in and out of equilibrium mentioned above, the limit $L_{x}\gg L$ has to be considered. This choice represents a sort of \textquotedblleft hadronic wire\textquotedblright : a cavity which is much longer in one spatial direction than in the other two (see \Cref{Figure_Cylinder}).
We will choose $\mathfrak{y}$ and $\mathfrak{z}$ as the \textquotedblleft homogeneous coordinates\textquotedblright\ (namely, the coordinates which do not appear explicitly in $T_{00}$ and $J^{0}$). When $L_{x}$ is not large compared to $L$, one must use the exact available results on SGT either on a finite interval or on $S^{1}$ developed in \cite{SGfinite1,SGfinite2} and references therein (we will come back to this case in a future publication).

As previously mentioned, the Skyrme field can be written explicitly in the Euler angles parametrization as in \Cref{I2}. A direct computation shows that, in terms of this parametrization, the Skyrme action is
\begin{align*}
I(H,F,G)=& -\frac{K}{2}\int d^{4}x\sqrt{-g}\biggl\{(\nabla H)^{2}+(\nabla F)^{2}+(\nabla G)^{2}+2\cos (2H)(\nabla F\cdot \nabla G) \\
& -\lambda \bigr(2\cos (2H)((\nabla H\cdot \nabla F)(\nabla H\cdot \nabla G)-(\nabla H)^{2}(\nabla F\cdot \nabla G)) \\
& +4\sin ^{2}(H)\cos ^{2}(H)((\nabla F\cdot \nabla G)^{2}-(\nabla F)^{2}(\nabla G)^{2}) \\
& +(\nabla H\cdot \nabla F)^{2}+(\nabla H\cdot \nabla G)^{2}-(\nabla H)^{2}(\nabla F)^{2}-(\nabla H)^{2}(\nabla G)^{2}\bigr)\biggl\}\ .
\end{align*}%
The energy-momentum tensor in this parametrization reads
\begin{equation*}
\begin{split}
T_{\mu \nu }=& -\frac{K}{2}\,\bigg\{-2[\nabla _{\mu }F\nabla _{\nu }F+\nabla_{\mu }H\nabla _{\nu }H+\nabla _{\mu }G\nabla _{\nu }G+\cos (2H)(\nabla
_{\mu }F\,\nabla _{\nu }G+\nabla _{\mu }G\,\nabla _{\nu }F)] \\
& +g_{\mu \nu }\,[(\nabla F)^{2}+(\nabla H)^{2}+(\nabla G)^{2}+2\cos(2H)(\nabla F\cdot \nabla G)] \\
& +2\lambda \biggl[\nabla _{\mu }F(\nabla H\cdot \nabla F)\nabla _{\nu}H-\nabla _{\mu }F(\nabla H)^{2}\nabla _{\nu }F-\nabla _{\mu }H(\nabla
F)^{2}\nabla _{\nu }H+\nabla _{\mu }H(\nabla F\cdot \nabla H)\nabla _{\nu }F \\
& +\nabla _{\mu }G(\nabla H\cdot \nabla G)\nabla _{\nu }H-\nabla _{\mu}G(\nabla H)^{2}\nabla _{\nu }G-\nabla _{\mu }H(\nabla G)^{2}\nabla _{\nu
}H+\nabla _{\mu }H(\nabla G\cdot \nabla H)\nabla _{\nu }G \\
& +\cos (2H)\,\big(\nabla _{\mu }F(\nabla H\cdot \nabla G)\nabla _{\nu}H-\nabla _{\mu }F(\nabla H)^{2}\nabla _{\nu }G -\nabla _{\mu }H(\nabla
F\cdot \nabla G)\nabla _{\nu }H \\
&+\nabla _{\mu }H(\nabla F\cdot \nabla H)\nabla _{\nu }G +\nabla _{\mu}G(\nabla H\cdot \nabla F)\nabla _{\nu }H-\nabla _{\mu }G(\nabla
H)^{2}\nabla _{\nu }F-\nabla _{\mu }H(\nabla G\cdot \nabla F)\nabla _{\nu }H \\
&+\nabla _{\mu }H(\nabla G\cdot \nabla H)\nabla _{\nu }F\big) +4\cos^{2}(H)\,\sin ^{2}(H)\,(\nabla _{\mu }F(\nabla G\cdot \nabla F)\nabla _{\nu
}G-\nabla _{\mu }F(\nabla G)^{2}\nabla _{\nu }F \\
&-\nabla _{\mu }G(\nabla F)^{2}\nabla _{\nu }G+\nabla _{\mu }G(\nabla F\cdot\nabla G)\nabla _{\nu }F)\biggl] \\
& +\lambda \,g_{\mu \nu }\,\big[(\nabla F)^{2}(\nabla H)^{2}-(\nabla F\cdot\nabla H)^{2}+(\nabla G)^{2}(\nabla H)^{2}-(\nabla G\cdot \nabla H)^{2}\\
& +2\cos (2H)\,((\nabla F\cdot \nabla G)(\nabla H)^{2}-(\nabla F\cdot \nabla H)(\nabla G\cdot \nabla H)) \\
& +4\cos ^{2}(H)\,\sin ^{2}(H)\,((\nabla F)^{2}(\nabla G)^{2}-(\nabla F\cdot\nabla G)^{2})\big]\bigg\}\;.
\end{split}
\end{equation*}
The field equations, obtained varying the action with respect to the degrees of freedom $F$, $H$ and $G$, are
\begin{align}  \label{fqomH}
0 &= \nabla_{\mu} \, \bigg\{\cos(2G)\,\sin(2H)\,\nabla^{\mu}F-\sin(2G)\,\nabla^{\mu}H  \notag \\
&- \lambda\,\sin(2G) \biggl((\nabla F)^{2} \nabla^{\mu}H - (\nabla F \cdot \nabla H) \nabla^{\mu}F + (\nabla G)^{2} \nabla^{\mu}H - (\nabla G \cdot
\nabla H)\nabla^{\mu}G  \notag \\
&+\cos(2H)\,( 2(\nabla F \cdot \nabla G) \nabla^{\mu} H - (\nabla F \cdot \nabla H) \nabla^{\mu} G - (\nabla H \cdot \nabla G) \nabla^\mu F)\biggl)
\notag \\
& - \lambda\,\cos(2G)\,\sin(2H)\biggl((\nabla F \cdot \nabla G) \nabla^\mu G+(\nabla F \cdot \nabla H) \nabla^\mu H - (\nabla G)^{2} \nabla^{\mu}F
\notag \\
&-(\nabla H)^{2} \nabla^{\mu} F - \cos(2H)\,\left((\nabla F \cdot \nabla G) \nabla^{\mu}F - (\nabla F)^{2}\nabla^{\mu}G\right)\biggl)\bigg\} \ , \\
0 &= \nabla_{\mu} \, \biggl\{\sin(2G)\,\sin(2H)\,\nabla^{\mu}F +\cos(2G)\,\nabla^{\mu}H  \notag \\
& + \lambda\,\cos(2G)\left( (\nabla G)^{2}\nabla^{\mu}H - (\nabla G \cdot \nabla H)\nabla^{\mu}G + (\nabla F)^{2} \nabla^{\mu}H - (\nabla F \cdot
\nabla H)\nabla^{\mu}F \right.  \notag \\
&\left.+ \cos(2H)\,\left(2(\nabla F \cdot \nabla G)\nabla^{\mu}H - (\nabla F \cdot \nabla H)\nabla^{\mu}G - (\nabla G \cdot \nabla H)\nabla^{\mu}F\right) \right)  \notag \\
& + \lambda\,\sin(2G)\,\sin(2H) ((\nabla H)^{2} \nabla^{\mu}F - (\nabla H \cdot \nabla F)\nabla^{\mu}H + (\nabla G)^{2} \nabla^{\mu}F  \notag \\
&-(\nabla G \cdot \nabla F)\nabla^{\mu}G + \cos(2H)((\nabla F \cdot \nabla G)\nabla^{\mu}F - (\nabla F)^{2} \nabla^{\mu}G))\biggl\} \ ,  \label{fqomF}
\\
0 &= \nabla_{\mu} \,\bigg\{\cos(2H)\,\nabla^{\mu}F + \nabla^{\mu}G - \lambda\,\sin^{2}(2H)\,((\nabla F \cdot \nabla G)\nabla^{\mu}F - (\nabla
F)^{2} \nabla^{\mu}G)  \notag \\
&+ \lambda\,\cos(2H)\,((\nabla H)^{2} \nabla^{\mu}F - (\nabla H \cdot \nabla F)\nabla^{\mu}H) + \lambda\,((\nabla H)^{2} \nabla^{\mu}G - (\nabla H \cdot \nabla G)\nabla^{\mu}H)\bigg\} \;.  \label{fqomG}
\end{align}
The only way to have an energy density homogeneous in $\mathfrak{y}$ and $\mathfrak{z}$ is to require that $F$ and $G$ are linear functions of these coordinates, and $H$ depends on the coordinate $x$ (transverse to the layer) and on time. The only Ansatz satisfying these properties is the one in \Cref{Ansatz01}. Indeed, if $H$ would depend either on $\mathfrak{y}$ or on $\mathfrak{z}$, then the energy density would depend on these coordinates as well. Hence, the profile $H$ carries the physical information on when and where $T_{00}$ and $J^{0}$ vanish and when they do not (that is why it makes sense to call $H$ \textquotedblleft profile\textquotedblright , as it encodes information on the spacetime variations of $T_{\mu \nu }$ and $J^{\mu }$). Furthermore, the above Ansatz is actually \textit{generic} if one is close enough to the boundary of any baryonic distribution (as it has been already emphasized). Consequently, \textit{the Ansatz here above describe locally any baryonic configuration close to one of its boundaries}.

The above choice has several remarkable properties (see \cite{56a0,56b1,gaugsksu(n),Canfora:2022jmh}). First, the three coupled nonlinear Skyrme field equations reduce consistently to just one PDE for the profile $H(t,x)$; the sine-Gordon equation in $(1+1)$ dimensions (as one
can check directly in \Cref{fqomH,fqomF,fqomG}). Second, this choice keeps alive the topological density.

In fact, by using the Ansatz in \Cref{Ansatz01}, the field equations in \Cref{fqomH,fqomF,fqomG}, are reduced to
\begin{equation}  \label{H_Skyrme_equation_second_derivative}
\partial_{t}^{2}H-\partial_{x}^{2}H+\frac{B^{2} \, \lambda}{8L^{2} \, (2L^{2} + B\,\lambda)}\sin(4H) = 0 \;,
\end{equation}
where we have considered for simplicity $p=q$, and $B=p^{2}>0$. Also, the on-shell Lagrangian density $\mathcal{L}_{on-shell}$ (apart from a constant term $-K\frac{B}{4L^{2}}$), the energy-density $T_{00}$ (apart from a constant term $K\frac{B}{4L^{2}}$), and the baryon density $\rho_{B}=J_{0}$ are, respectively,
\begin{equation}
\mathcal{L}_{on-shell} = \frac{K}{64L^{4}} \, \left(16L^{2} \, \left(2L^{2} + \left\vert B \right\vert \, \lambda\right) \, \left[\left(\partial_{t}H\right)^{2} - \left(\partial_{x}H\right)^{2}\right] - B^{2} \, \lambda \, \left(1 - \cos\left(4H\right)\right)\right) \;,
\end{equation}
\begin{equation}
T_{00} = \frac{K}{64L^{4}} \, \left(16L^{2}\left(2L^{2} + \left\vert B \right\vert \, \lambda\right) \, \left[\left(\partial_{t}H \right)^{2} + \left(\partial_{x}H\right)^{2}\right] + B^{2} \, \lambda \left(1-\cos(4H)\right)\right) \;,
\end{equation}
\begin{equation}  \label{SGE1}
\ \rho_{B} = J_{0} = -3 \, B \, \left(\partial_{x}H\right) \, \sin(2H)\;.
\end{equation}
The boundary conditions for $H$ are fixed by requiring that the baryonic charge is $\pm B$, as we have mentioned. For the case $Q_{B}$ to be zero, should be $\cos(2H(t,L_{x}))=\cos(2H(t,0))$, $\forall\,t$. This implies
\begin{equation}  \label{BC3}
H(t,L_{x})=H(t,0) + n\,\pi, \quad \mbox{\emph{or}} \quad H(t,L_{x}) + H(t,0)= m\,\pi, \quad n,m\in\mathds{Z}, \quad \forall\,t \;.
\end{equation}
Such configurations, of $Q_{B}=0$, are interesting anyway since the baryonic density is nontrivial and one can have bound states of two layers (in breatherlike style).

For static configurations, $H(t,x)=H(x)$, one can reduce the field equation to a simple quadrature
\begin{equation}
\left( \partial _{x}H\right) =\pm \left[ E_{0}-\frac{B^{2}\,\lambda }{16L^{2}\left( 2L^{2}+B\,\lambda \right) }\cos 4H\right] ^{1/2}\ ,
\label{H_Skyrme_equation_first_derivative}
\end{equation}
where the integration constant $E_{0}$ is fixed by
\begin{equation}
4\int_{0}^{\pi /2}\frac{dH}{\left[ 16L^{2}E_{0}-\frac{B^{2}\lambda }{\left(
2L^{2}+B\,\lambda \right) }\cos 4H\right] ^{1/2}}=\frac{L_{x}}{L}\;,
\label{BC4}
\end{equation}%
where we have taken, for simplicity, $n=m=0$ in \Cref{BC3}. It is easy to see that the above equation for $E_{0}$ always has a solution if $L_{x}$ is finite. Indeed, if $L_{x}$ is small (compared to $L$) one can take a large $E_{0}$ to make the left-hand side of \Cref{BC4} small as well. If $L_{x}$ is large (but not divergent), one can have the left-hand side of \Cref{BC4} large by choosing
\begin{equation*}
E_{0}=\frac{B^{2}\lambda }{16L^{2}\left( 2L^{2}+B\lambda \right)}+\varepsilon \ ,\ \ 0<\varepsilon \ll 1\ ,
\end{equation*}%
so that the denominator of the left-hand side of \Cref{BC4} comes close to have a zero when $H=0$, or $H=\pi /2$. The $L_{x}\rightarrow \infty$ case corresponds to the limit in which $\varepsilon =0$.

The proper normalization of the Skyrme profile $H$ to define the effective sine-Gordon Lagrangian $L_{SG}$ and the corresponding energy-density $\widetilde{T}_{00}$ can be achieved by requiring that the integral along the coordinate $x$ of $\widetilde{T}_{00}$ \textit{should give the actual total energy of the hadronic layers} (with a similar condition for the effective action $I_{SG}$). Hence, $\widetilde{T}_{00}$ is the integral in the transverse coordinates $\mathfrak{y}$ and $\mathfrak{z}$ of $T_{00}$, so that the integral along $x$ of $\widetilde{T}_{00}$ will give the total energy of the Skyrmionic system
\begin{equation}
\widetilde{T}_{00}=\frac{\pi ^{2}K}{8L^{2}}\left\{ 16L^{2}\left(2L^{2}+\left\vert B\right\vert \lambda \right) \,\left[ \left( \partial_{t}H\right) ^{2}+\left( \partial _{x}H\right) ^{2}\right] +B^{2}\,\lambda \left( 1-\cos (4H)\right) \right\} \;,  \label{effectivesg1}
\end{equation}%
where the factor $8\pi ^{2}L^{2}$ comes from the integral in $\mathfrak{y}$ and $\mathfrak{z}$. The same is true for the effective Lagrangian $L_{SG}$:
\begin{equation}
L_{SG}=\frac{\pi ^{2}K}{8L^{2}}\left\{ 16L^{2}\left( 2L^{2}+B\,\lambda\right) \left[ \left( \partial _{t}H\right) ^{2}-\left( \partial_{x}H\right) ^{2}\right] -B^{2}\lambda \left( 1-\cos \left( 4H\right) \right) \right\} \;.  \label{effectivesg2}
\end{equation}
The proper normalization of the kinetic term can be achieved normalizing $H$ as follows:
\begin{equation}
\varphi =\frac{4}{\beta }H\ ,\quad \beta =\frac{2}{\pi \left[ K\left(2L^{2}+B\,\lambda \right) \right] ^{1/2}}\ ,\qquad M_{0}=\frac{\pi ^{2}K}{8L^{2}}B^{2}\lambda \;,
\end{equation}
so that, the effective sine-Gordon coupling $\beta$ and the effective dimensionless sine-Gordon action become, respectively,
\begin{equation}
\beta =\frac{2}{\pi \left[ K\left( 2L^{2}+B\,\lambda \right) \right] ^{1/2}} \;,
\end{equation}
\begin{equation}
I_{SG}=\int L_{SG}dtdx=\,\,\int \left( -\frac{1}{2}\partial ^{\mu }\varphi\partial _{\mu }\varphi +M_{0}\left( \cos (\beta \,\varphi )-1\right) \right) dt\,dx \;,
\end{equation}
where constant terms have been discarded.

\subsection{Perturbations on the solutions}
\label{section_technical_perturbations}

An important technical part of the present work is to show that, with the Ansatz defined in \Cref{Ansatz01}, not only the field equations and the energy-density reduce to the corresponding quantities in SGT in $(1+1)$ dimensions in a sector with nonvanishing baryonic charge, but also that the lowest energy perturbations of these configurations are precisely perturbations of the sine-Gordon effective field, which only depend on $t$ and $x$ (this is the reason why it is convenient to take $L_{x}\gg L$: in this case, the energy needed to excite modes that depend nontrivially on the transverse coordinates is much higher than the energy needed to excite ``sine-Gordon modes"). This issue is relevant since, if we want to use the available results on equilibrium and nonequilibrium SGT (in particular, \cite{SG1a,SG1b,SG2,SG3a,SG3b,SG3c,SG4a,SG4b,SG4c,SG5,SG6,Koch:2023xuw,SG7,SGfinite1,SGfinite2,entanglement9aa1,entanglement9aa2}), then we must identify a regime in which also the low energy fluctuations are of ``sine-Gordon type''.

Hence, let us consider a general solution $U_{0}$ of the form in \Cref{I2}, where $F$, $H$ and $G$ fulfilling \Cref{fqomH,fqomF,fqomG}. Let us now take a perturbation of $U_{0}$ as
\begin{equation}
U\ =U_{0}\left( \mathbb{1}+\chi \right) \;,  \label{Upert}
\end{equation}%
where $\chi $ is a $2\times 2$ matrix with the conditions
\begin{equation}
\chi ^{\dagger }=-\chi \;,\qquad \mbox{Tr}\,\chi =0\,.
\label{chi_conditions}
\end{equation}%
These conditions ensure that $\chi $ is an arbitrary element of the $\mathfrak{su}(2)$ algebra, i.e.,
\begin{equation}
\chi =\epsilon \,\chi _{1}\,t_{1}\,+\epsilon \,\chi _{2}\,t_{1}\,+\epsilon
\,\chi _{3}\,t_{3}\;,  \label{chi}
\end{equation}%
where $\left\vert \epsilon \right\vert \ll 1$ is the perturbation parameter. Observe that $\chi _{i}$ are real functions of the coordinates $t$, $x$, $\mathfrak{y}$ and $\mathfrak{z}$.\footnote{Notice that we could have equally defined $U\ =U_{0}+\chi ^{\prime }$, where $\chi^{\prime}=U_{0}\chi $. Therefore, as $U_{0}$ is invertible, all the results for $\chi $ are equivalent to $\chi^{\prime}$. In this way, we see that \Cref{Upert} is directly related to the usual prescription to write a perturbation as the solution $U_{0}$ plus \textit{something}.} Introducing this expansion in \Cref{Eqoskyrme}, we get
\begin{align}
\lbrack R_{0\mu }+\frac{\lambda }{4}[R_{0}^{\nu },G_{0\mu \nu }],\nabla^{\mu }\chi ]+\nabla ^{\mu }\bigg(\nabla _{\mu }\chi +\frac{\lambda }{4}[\nabla ^{\nu }\chi ,G_{0\mu \nu }]& \bigg)  \notag  \label{field_eq_chi} \\
+\frac{\lambda }{4}\nabla ^{\mu }\bigg(\lbrack R_{0}^{\nu },[\nabla _{\mu}\chi ,R_{0\nu }]]+[R_{0}^{\nu },[R_{0\mu },\nabla _{\nu }\chi ]]& \bigg)=0\;,
\end{align}
up to order $O(\epsilon ^{2})$. Now, let us consider a generic perturbation (in the present context, the wording ``generic perturbation'' means that we allow the perturbation to depend on all four space-time coordinates). A natural Ansatz for the perturbation is
\begin{align*}
\chi ^{1}\ =\ & \zeta _{1}(x)\,\cos (H(x))\,\cos (p\,\mathfrak{z})\,e^{\mathrm{i}(\omega \,t+k_{1}\,\mathfrak{y}+k_{2}\,\mathfrak{z})}\;, \\
\chi ^{2}\ =\ & \zeta _{2}(x)\,\cos (H(x))\,\sin (p\,\mathfrak{z})\,e^{\mathrm{i}(\omega \,t+k_{1}\,\mathfrak{y}+k_{2}\,\mathfrak{z})}\;, \\
\chi ^{3}\ =\ & \zeta _{3}(x)\,\sin (H(x))\,e^{\mathrm{i}(\omega \,t+k_{1}\,\mathfrak{y}+k_{2}\,\mathfrak{z})}\;, \qquad k_{1}\neq 0\ ,\ k_{2}\neq 0 \;,
\end{align*}
where we are taking into account the fact that the energy and baryon densities do not depend on the transverse coordinates. The profiles $\zeta_{j}(x)$\ ($j=1,2,3$) of the perturbations only depend on $x$. The first conclusion which arises analyzing the linearized field equations is that, actually, only one of these three profiles is independent. Namely, one can choose\footnote{The fact that only one radial function is necessary in the Ansatz can be easily verified for the nonlinear sigma model case, that is, when $\lambda= 0$. Moreover, for small values of $\lambda$ this could still be true by analytic continuation. It can also be checked that the linearized equations for perturbations of the form in \Cref{Ansatz_perturbation} are always consistent (namely, for any value of $\lambda$ one always gets as many equations as unknown functions).}: $\zeta _{1}(x)=\zeta _{2}(x)=\zeta (x)=-\zeta _{3}(x)$. Now, considering
\begin{align}
\chi ^{1}\ =\ & \zeta (x)\,\cos (H(x))\,\cos (p\,\mathfrak{z})\,e^{\mathrm{i}(\omega \,t+k_{1}\,\mathfrak{y}+k_{2}\,\mathfrak{z})}\;,  \notag
\label{Ansatz_perturbation} \\
\chi ^{2}\ =\ & \zeta (x)\,\cos (H(x))\,\sin (p\,\mathfrak{z})\,e^{\mathrm{i}(\omega \,t+k_{1}\,\mathfrak{y}+k_{2}\,\mathfrak{z})}\;, \\
\chi ^{3}\ =\ & -\zeta (x)\,\sin (H(x))\,e^{\mathrm{i}(\omega \,t+k_{1}\,\mathfrak{y}+k_{2}\,\mathfrak{z})}\;,  \notag
\end{align}
one can check that the complete set of linearized Skyrme equations (\ref{field_eq_chi}) are satisfied if $k_{2}=\frac{q}{p}\,k_{1},$ (i.e., $k_{2}=k_{1}\equiv k$, because we are taken $p=q$), and if $\zeta $ satisfies a linear ordinary differential equation (ODE) of the form $\zeta^{\prime\prime }(x)+A(x)\,\zeta ^{\prime }(x)+B(x)\,\zeta (x)\ =\ 0$, (where the functions $A(x)$ and $B(x)$ can be computed explicitly in terms of the background solution). In order to use the Sturm-Liouville theory, it is convenient the following change of variables: $\zeta(x)=\alpha(x)\,\xi (x)$, choosing $\alpha $ in such a way to eliminate the first derivative term. In this way, we get
\begin{equation}
-\xi ^{\prime \prime }(x)+\frac{Q(x)}{L^{2}}\,\xi (x)=\omega ^{2}\,W(x)\,\xi(x)\;,  \label{pre_Sturm-Liouville}
\end{equation}
where the functions $Q(x)$ and $W(x)$ are
\begin{align*}
Q(x)=& \frac{1}{32\,(B\uplambda+2)\,(B\,\uplambda\,\cos ^{2}(H(x))+2)^{2}}\,\times \\
& \bigg\{(B\uplambda\cos (2H(x))+B\uplambda+4)\,\bigg(B^{3}\,\uplambda^{2}\,\cos (6H(x)) +B^{2}\uplambda\,(3B\,\uplambda+8)\,\cos (2H(x)) \\
&+2(B\uplambda+2)\,(B^{2}\,\uplambda\,\cos (4H(x))+B^{2}\,\uplambda+32k^{2})\bigg) \\
& -\bigg(B\,\uplambda\bigg(4\cos (2H(x))\,(\uplambda(B-k^{2})+3)+B\,\uplambda\,\cos (4H(x))\bigg) \\
& +B\,\uplambda^{2}\,(3B-4k^{2})+8\uplambda\,(B-2k^{2})+8\bigg)\,\bigg(16(B\,\uplambda+2)\,E_{0}+B^{2}\,\uplambda\,\cos (4H(x))\bigg)\bigg\}\;, \\
W(x)=& \frac{4\uplambda\,E_{0}+B\,\uplambda\left( \frac{B\uplambda\,\cos(4H(x))}{4B\,\uplambda+8}+\cos (2H(x))\right) +B\,\uplambda+4}{(2B\,\uplambda\,\cos ^{2}(H(x))+4)}\;.
\end{align*}
Here, we have defined $\uplambda\equiv \frac{\lambda }{L^{2}}$, and used \Cref{H_Skyrme_equation_second_derivative,H_Skyrme_equation_first_derivative}. Also,
\begin{equation}
\alpha (x)=\frac{C_{1}}{\sqrt{4+B\,\uplambda(1+\cos (H(x)))}}\;,
\label{alpha}
\end{equation}%
being $C_{1}$ an arbitrary dimensionless constant. \Cref{pre_Sturm-Liouville} can be written in a slightly different manner by the change of variable $\rho =\frac{x}{L}$,
\begin{equation}
-\xi ^{\prime \prime }+Q\,\xi =\upomega^{2}\,W\,\xi \;,
\label{Sturm-Liouville}
\end{equation}
where the substitution of $x$ as a function of $\rho $ is carried out whatever is necessary, the prime now denotes $\frac{\partial}{\partial\rho}$, and we have defined $\upomega\equiv \omega \,L$.

\begin{figure}[tbp]
\begin{subfigure}[b]{0.32\textwidth}
\includegraphics[width=\textwidth]{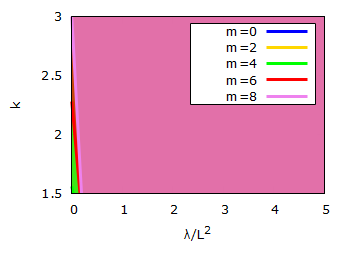} %
\caption{$B=0$ and $\frac{L_{x}}{L}=1$}
\end{subfigure}
\hfill
\begin{subfigure}[b]{0.32\textwidth}
\includegraphics[width=\textwidth]{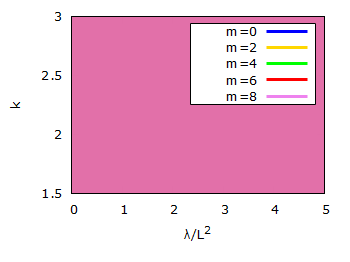}
\caption{$B=0$ and $\frac{L_{x}}{L}=5$}
\end{subfigure}
\hfill
\begin{subfigure}[b]{0.32\textwidth}
\includegraphics[width=\textwidth]{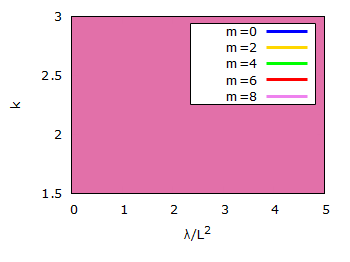}
\caption{$B=0$ and $\frac{L_{x}}{L}=10$}
\end{subfigure}
\hfill
\begin{subfigure}[b]{0.32\textwidth}
\includegraphics[width=\textwidth]{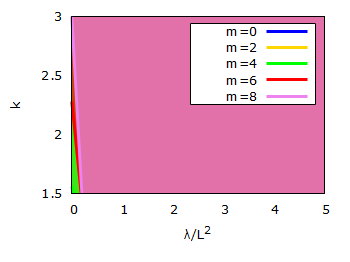}
\caption{$B=1$ and $\frac{L_{x}}{L}=1$}
\end{subfigure}
\hfill
\begin{subfigure}[b]{0.32\textwidth}
\includegraphics[width=\textwidth]{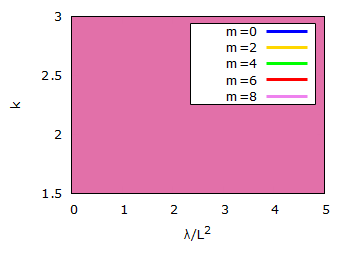}
\caption{$B=1$ and $\frac{L_{x}}{L}=5$}
\end{subfigure}
\hfill
\begin{subfigure}[b]{0.32\textwidth}
\includegraphics[width=\textwidth]{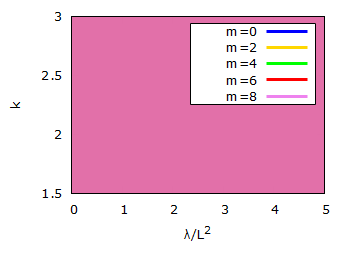}
\caption{$B=1$ and $\frac{L_{x}}{L}=10$}
\end{subfigure}
\hfill
\begin{subfigure}[b]{0.3\textwidth}
\includegraphics[width=\textwidth]{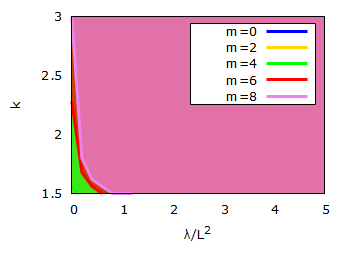}
\caption{$B=4$ and $\frac{L_{x}}{L}=1$}
\end{subfigure}
\hfill
\begin{subfigure}[b]{0.32\textwidth}
\includegraphics[width=\textwidth]{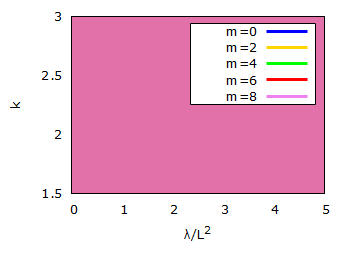}
\caption{$B=4$ and $\frac{L_{x}}{L}=5$}
\end{subfigure}
\hfill
\begin{subfigure}[b]{0.32\textwidth}
\includegraphics[width=\textwidth]{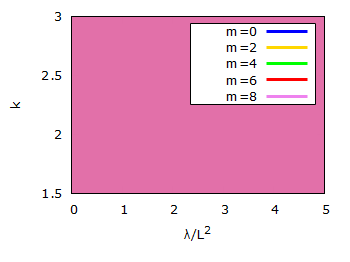}
\caption{$B=4$ and $\frac{L_{x}}{L}=10$}
\end{subfigure}
\hfill
\begin{subfigure}[b]{0.32\textwidth}
\includegraphics[width=\textwidth]{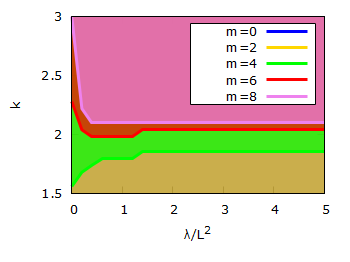}
\caption{$B=9$ and $\frac{L_{x}}{L}=1$}
\end{subfigure}
\hfill
\begin{subfigure}[b]{0.32\textwidth}
\includegraphics[width=\textwidth]{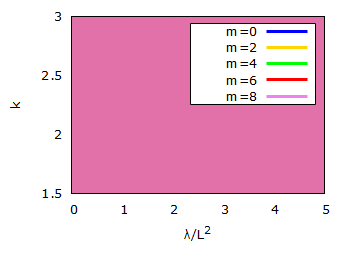}
\caption{$B=9$ and $\frac{L_{x}}{L}=5$}
\end{subfigure}
\hfill
\begin{subfigure}[b]{0.32\textwidth}
\includegraphics[width=\textwidth]{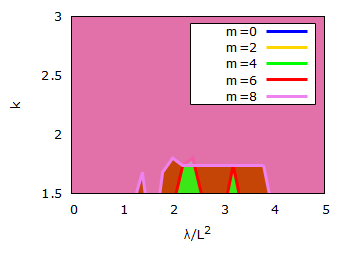}
\caption{$B=9$ and $\frac{L_{x}}{L}=10$}
\end{subfigure}
\caption{Plots of the stability regions at the $(k\,\protect\lambda /L^{2})$ plane for $m=0,2,4,6,8$ in (\protect\ref{BC1}) for different values of $B$ and $\frac{L_{x}}{L}$. For example, plot (j) shows that above the separation line of $m=8$ (in violet) the region is stable, while below such separation line is unstable; for $m=6$, above the separation line (in red) is stable, while below it is unstable; same for the separation line $m=4$ (in green) and $m=2$ (in yellow). Plot (b) shows that, for those values of $\frac{L_{x}}{L}$ and $B$, the region is stable for all values of $m\leq8$. Notice that the $k$ values are representative, as by boundary condition in the cavity, the allowed values of $k$ are, in fact, integers.}
\label{Figure_k_vs_lambdaL2}
\end{figure}

\begin{figure}[tbp]
\begin{subfigure}{0.45\textwidth}
\includegraphics[angle=0,width=\textwidth]{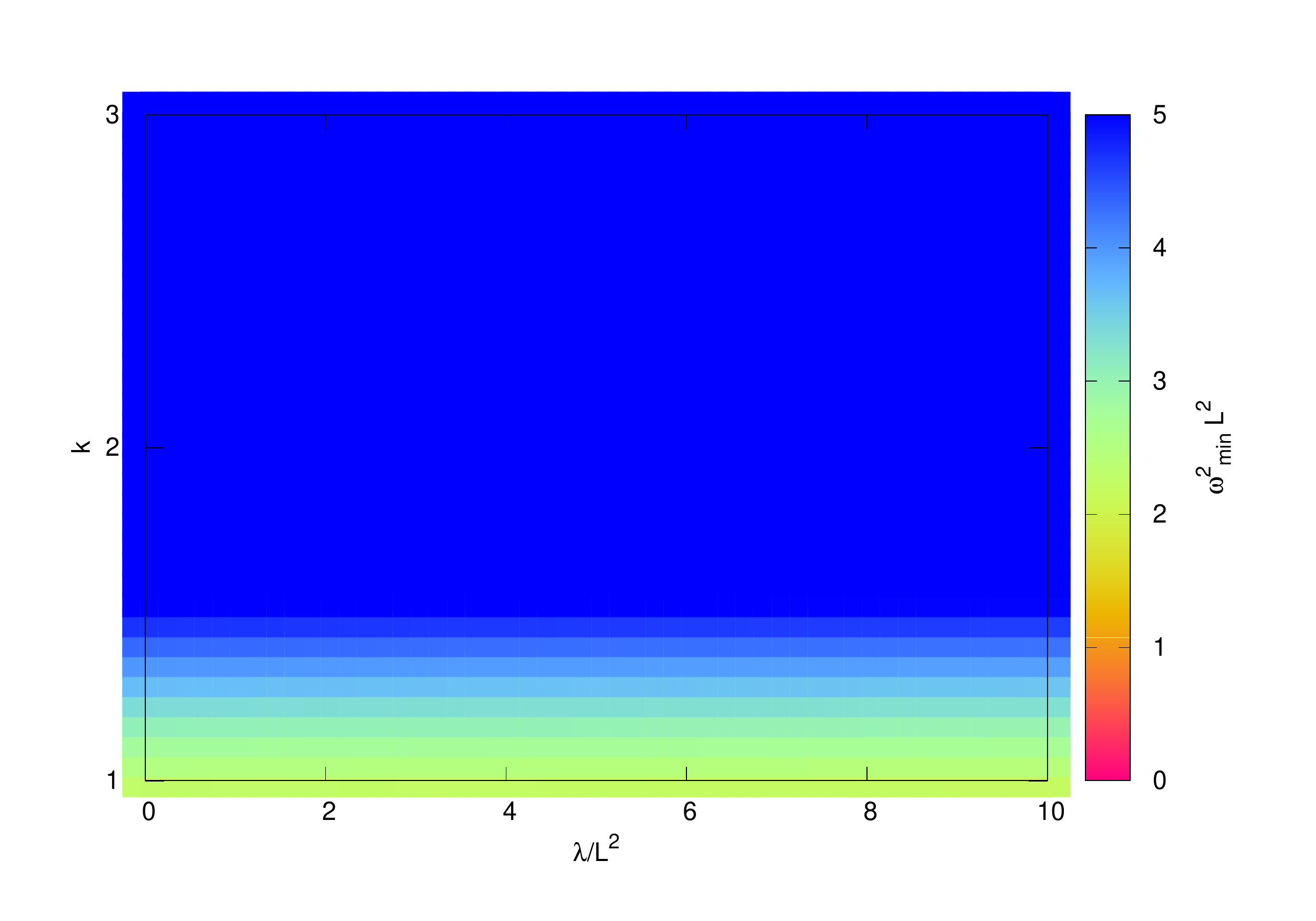}
\caption{$B=0$ and $\frac{L_{x}}{L}=1$}
\end{subfigure}
\hfill
\begin{subfigure}{0.45\textwidth}
\includegraphics[angle=0,width=\textwidth]{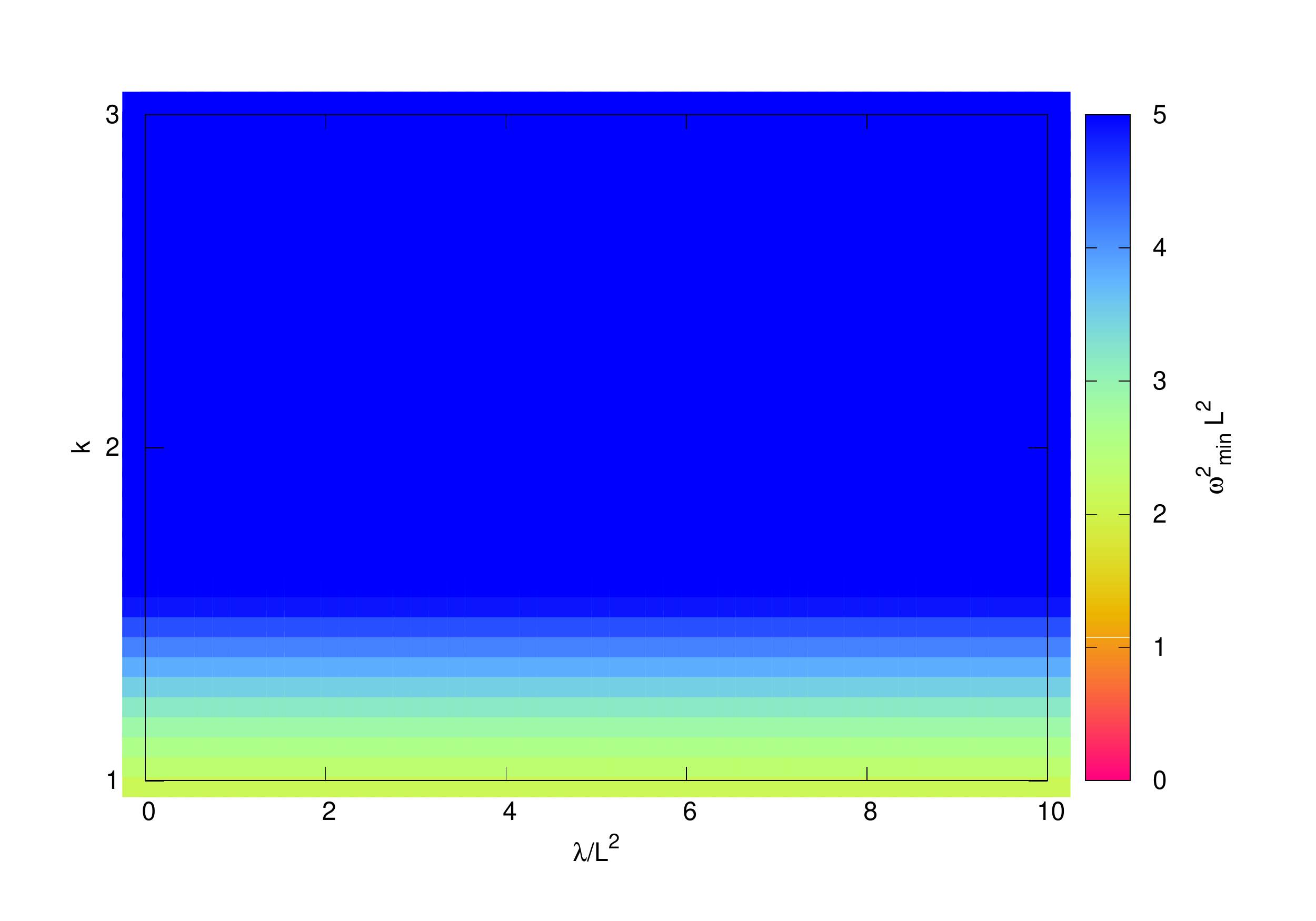}
\caption{$B=0$ and $\frac{L_{x}}{L}=5$}
\end{subfigure}
\hfill
\begin{subfigure}{0.45\textwidth}
\includegraphics[angle=0,width=\textwidth]{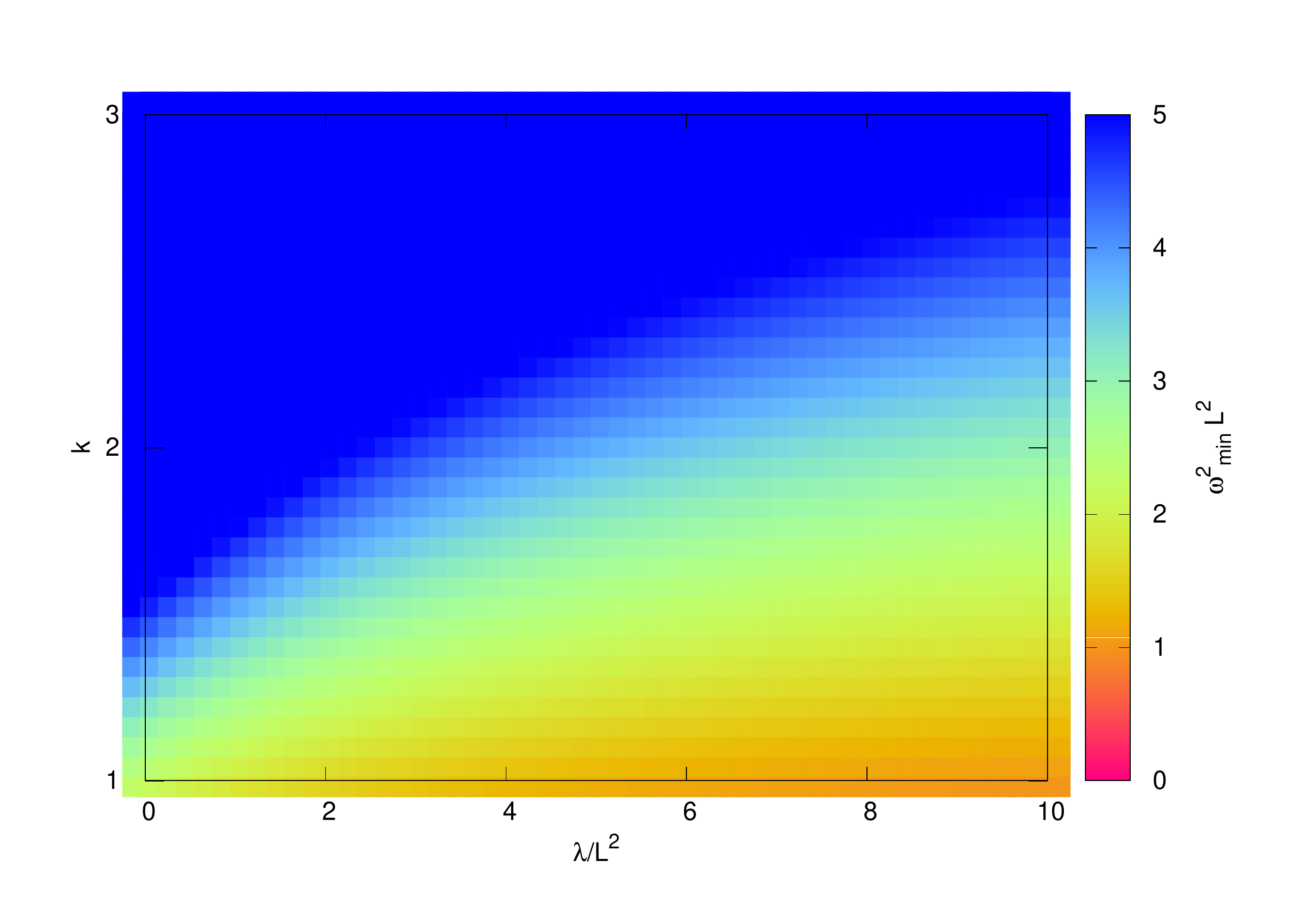}
\caption{$B=1$ and $\frac{L_{x}}{L}=1$}
\end{subfigure}
\hfill
\begin{subfigure}{0.45\textwidth}
\includegraphics[angle=0,width=\textwidth]{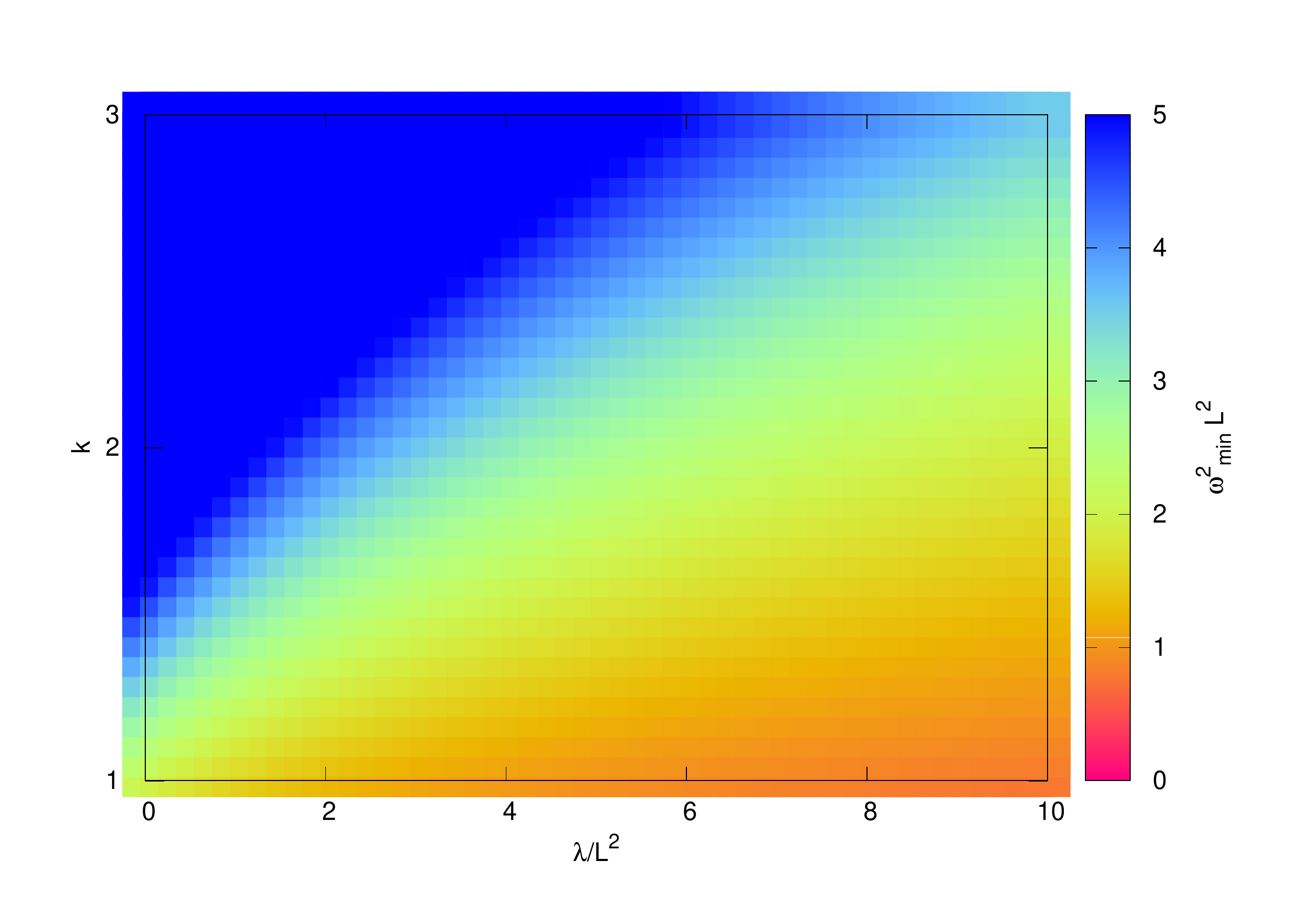}
\caption{$B=1$ and $\frac{L_{x}}{L}=5$}
\end{subfigure}
\hfill
\begin{subfigure}{0.45\textwidth}
\includegraphics[angle=0,width=\textwidth]{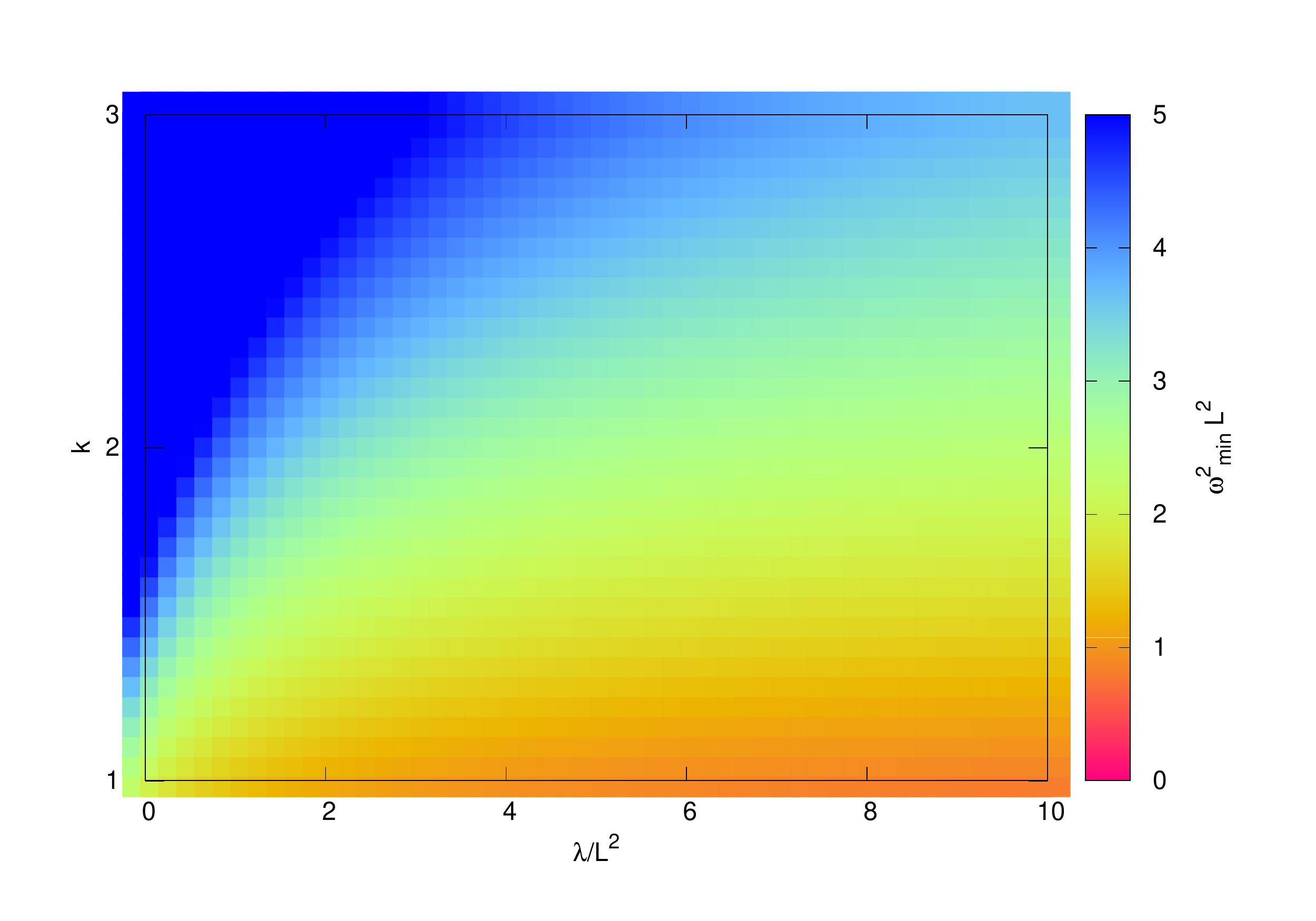}
\caption{$B=4$ and $\frac{L_{x}}{L}=1$}
\end{subfigure}
\hfill
\begin{subfigure}{0.45\textwidth}
\includegraphics[angle=0,width=\textwidth]{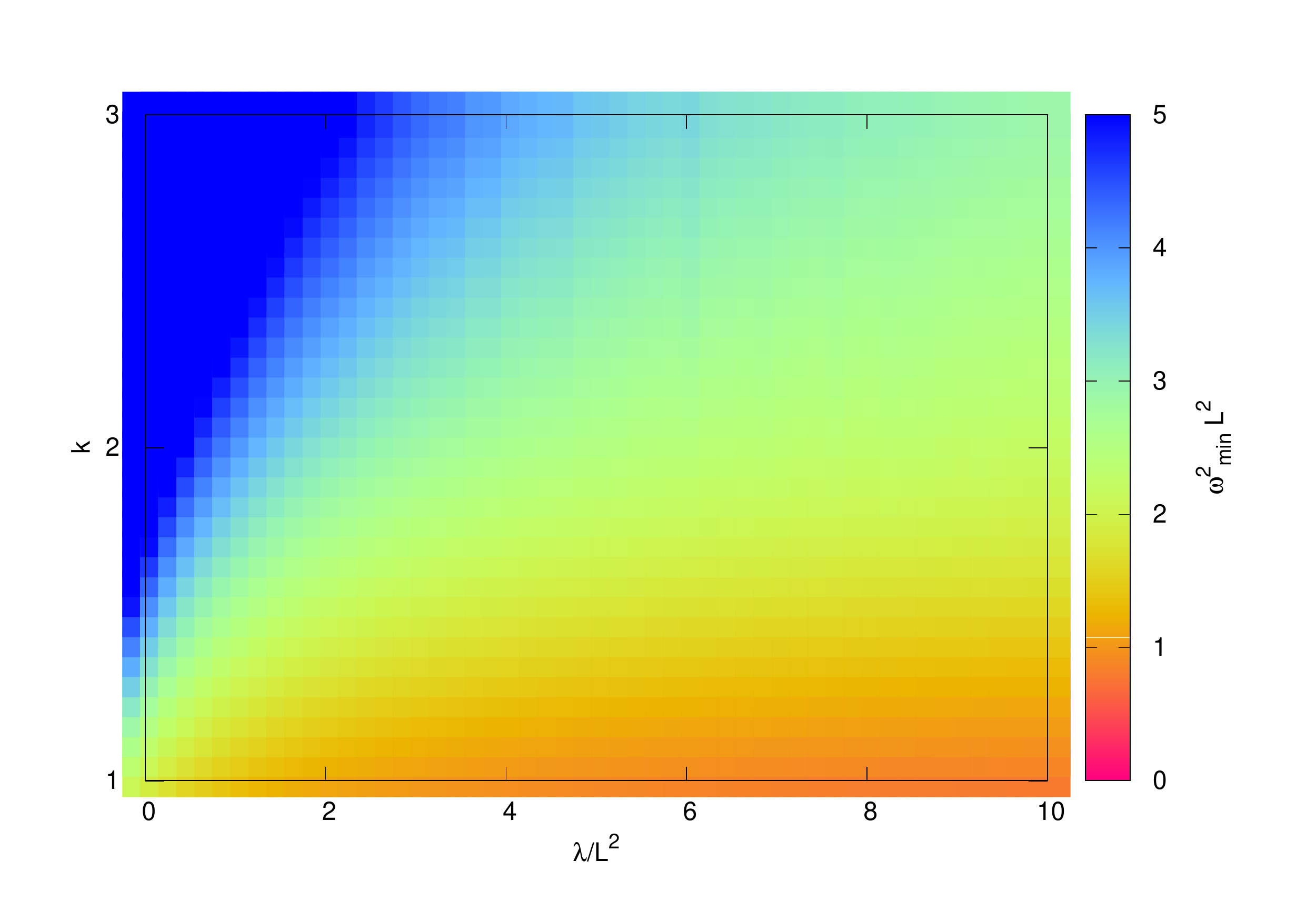}
\caption{$B=4$ and $\frac{L_{x}}{L}=5$}
\end{subfigure}
\hfill
\begin{subfigure}{0.45\textwidth}
\includegraphics[angle=0,width=\textwidth]{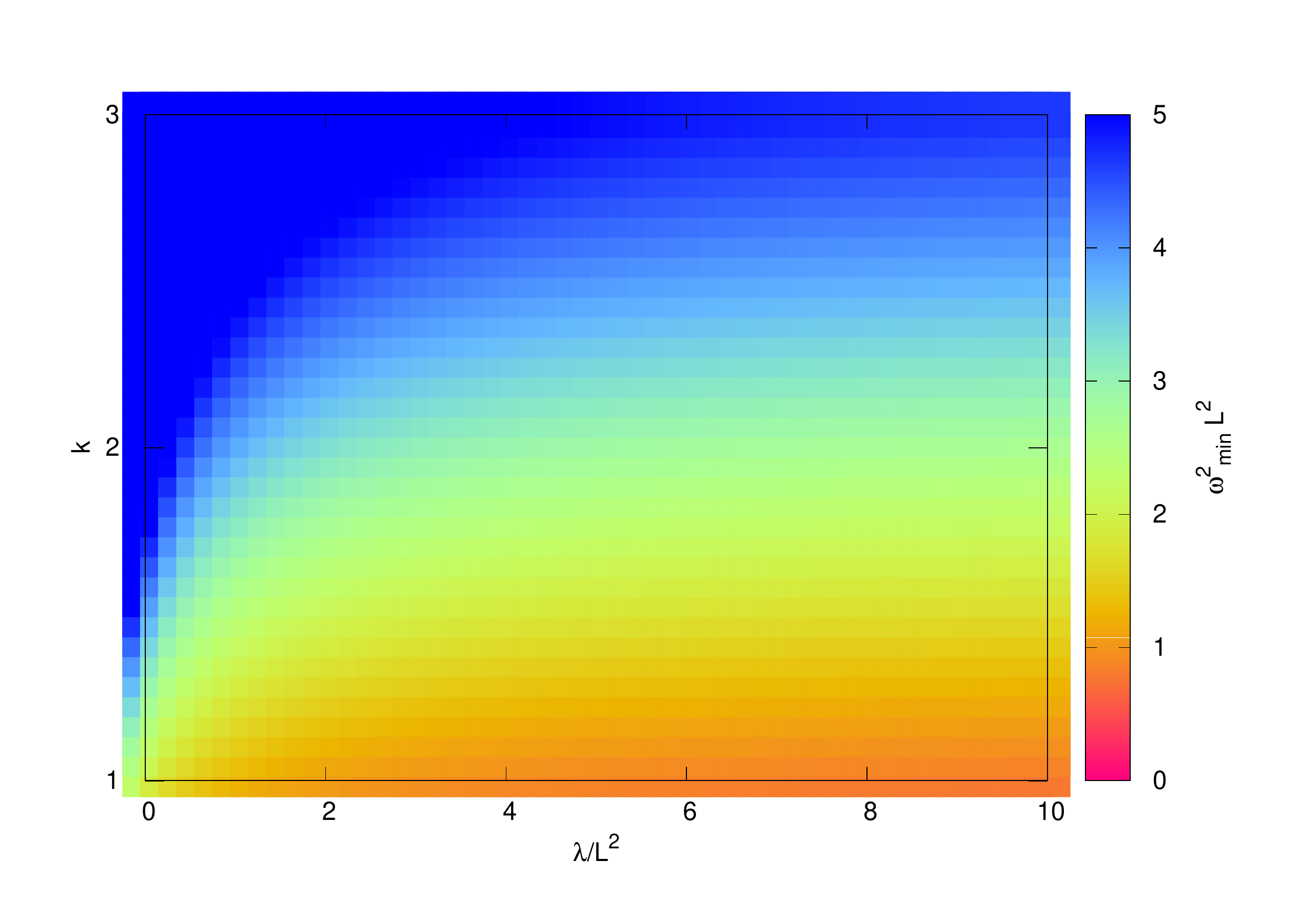}
\caption{$B=9$ and $\frac{L_{x}}{L}=1$}
\end{subfigure}
\hfill
\begin{subfigure}{0.45\textwidth}
\includegraphics[angle=0,width=\textwidth]{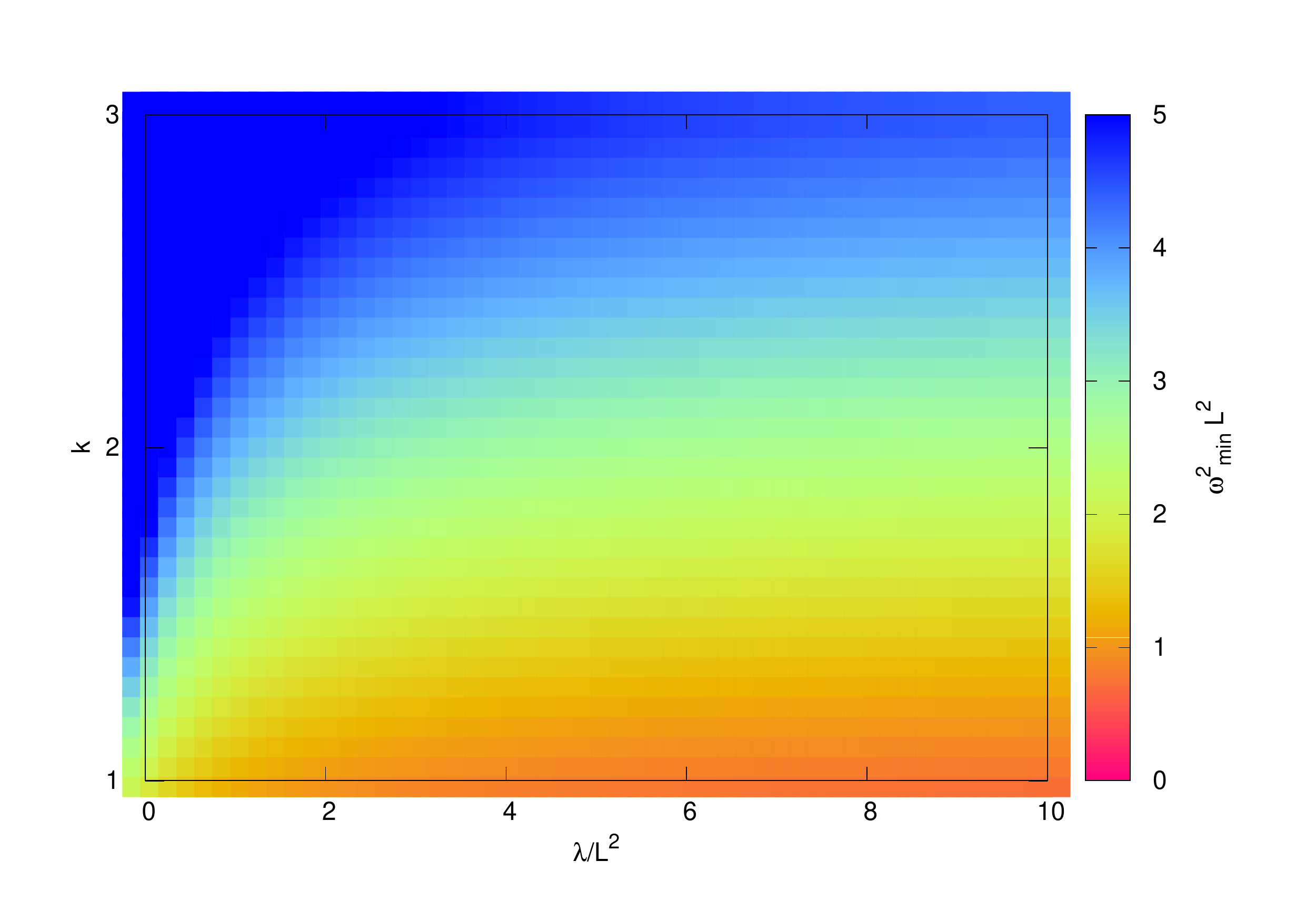}
\caption{$B=9$ and $\frac{L_{x}}{L}=5$}
\end{subfigure}
\hfill 
\caption{Plots $k$ vs. $\protect\lambda/L^{2}$ for $m=0$ taking different values of $B$ and $\frac{L_{x}}{L}$. The dimensionless quantities $\upomega^{2}_{\mbox{min}}=\protect\omega^{2}_{\mbox{min}}\,L^{2}$ are shown in the color palette on the right of each plot. As the values of $\upomega^{2}$ are not much tiny, $\protect\omega$ goes as $L^{-1}$, at least for the set of parameters analyzed. The $k$ values are representative (see the caption of Figure \protect\ref{Figure_k_vs_lambdaL2}).}
\label{Figure_omega2min}
\end{figure}

The ODE in \Cref{Sturm-Liouville} is a particular case of the Sturm-Liouville problem \footnote{Usually, the SLP is written as $-\xi ^{\prime\prime }+q\,\xi=\lambda\,W\,\xi $, being $\lambda $ the eigenvalues to be found for particular boundary conditions \cite{Pryce1993}. Therefore, \Cref{Sturm-Liouville}) is a particular case with $p=1$, $q=Q$, $w=W$, and eigenvalues $\lambda =\upomega^{2}$.} (SLP), it can hardly be solved analytically and even numerically, it is not simple. For us it is enough to have some sufficient stability conditions: we require that the function $\xi$ does not diverge inside the range where $x$ is defined, i.e., the interval $[0,L_{x}]$. Therefore, we find the eigenvalues $\upomega^{2}$ using the method of \emph{simple centred differences} \cite{Pryce1993}, and the minimum $\upomega^{2}_{\mbox{min}}$ should be positive for stability, with boundary conditions $\xi (0)=0$ and $\xi (L_{x})=0$. The function $H(x)$ as a solution of \Cref{H_Skyrme_equation_second_derivative}, with boundary conditions in \Cref{BC1}, is well-behaved. Also, the function in \Cref{alpha} is well-behaved, strictly positive and without singularities inside the integration range if $\lambda \geq 0$ and $L\neq 0$. In \Cref{Figure_k_vs_lambdaL2}, we show the stability regions for different values of $B$ and $\frac{L_{x}}{L}$ (see details in the caption).
By analyzing the energy scale of the fluctuations of $F$, $G$ and $H$, it is observed that the minimum of positive frequencies $\omega_{\mbox{min}}$ goes as $1/L$ due to the scale normalization of $\upomega$ in \Cref{Sturm-Liouville}, at least for the set of parameters analyzed. This can be seen in \Cref{Figure_omega2min} for some values of $B$ and $\frac{L_{x}}{L}$.
On the other hand, the lowest energy perturbations are small variations $\delta H$ of $H(t,x)$ which depend only on $t$ and $x$. In particular, these perturbations of static profile $H(t,x)=H(x)$ are gapless, as in SGT (when $L_{x}\gg L$). One can readily see this as follows: The Skyrme field equations for a static profile $H(t,x)=H(x)$ corresponding to the Ansatz in \Cref{Ansatz01} reduces to \Cref{H_Skyrme_equation_second_derivative}, which, in its turn, reduces to \Cref{H_Skyrme_equation_first_derivative}, where the condition in \Cref{BC4} fixes the integration constant $E_{0}$. Given a solution $H_{0}(x)$ of \Cref{H_Skyrme_equation_first_derivative} one can always find a solution $\delta H$ of the linearized field equation with zero energy as $\delta H=\partial _{x}H_{0}(x)$. With the appropriate choice of $E_{0}$, $\partial _{x}H_{0}(x)$ never changes sign, so $\partial_{x}H_{0}(x)$ is a nodeless zero mode. Moreover, as has been shown in Ref. \cite{Takayama1993}, SGT possesses gapless modes.

Summarizing, the above arguments show that at energy and/or temperatures less than $1/L$, the only modes that are energetically available in the full Skyrme theory in the cavity, like \Cref{Figure_Cylinder}, are the sine-Gordon modes associated with perturbations $\delta H$ of $H(t,x)$ which depend only on $t$ and $x$. Consequently, not only does the Skyrme model reduce to SGT for baryonic layers configurations in such a cavity, but also perturbations in this regime are, in fact, the lowest energy perturbations of SGT because generic perturbations like the ones shown above always possess higher energy.


\section{Conclusions}


In this article, we have constructed an exact mapping between the Skyrme model in $3+1$ dimensions at finite Baryon density and the sine-Gordon model in $(1+1)$ dimensions. Such mapping is valid for a distribution of baryonic matter close to its boundaries (as explained in the previous sections) and for low enough energy. This mapping opens a new window to analyze many equilibrium and nonequilibrium phenomena of hadronic matter that can be fully understood neither with perturbation theory nor with lattice QCD using well-known results in SGT.  These analytic results (especially the out-of-equilibrium ones) are entirely out of reach of the other available theoretical methods in the low-energy sector of QCD. As examples, we have discussed the robust predictions on the oscillations of von Neumann and R{\'e}nyi entropies. Still, it is expected that the present results will generate many more surprises that are difficult to envisage right now: the physical consequences of this mapping are far-reaching. They will be further investigated in forthcoming papers.

Finally, it is worth mentioning that the present results are also intriguing in the analysis of neutron stars. Indeed, configurations such as hadronic tubes and layers of baryons (known as nuclear pasta states) appear \cite{Newton2013,Pons:2013nea}. Our framework allows us to determine, among other things, the transport properties of these inhomogeneous baryonic distributions with such beautiful shapes, which are challenging to compute using numerical simulations.


\section*{Acknowledgements}


F.~C. has been funded by Fondecyt Grant No.~1200022. M.~L. is funded by ANID, Convocatoria Nacional Subvención a la Instalación en la Academia Convocatoria Año 2022, Folio SA85220027. P.~P. is supported by Fondo Nacional de Desarrollo Cient\'{\i}fico y Tecnol\'{o}gico--Chile (Fondecyt Grant No.~3200725) and by Charles University Research Center (UNCE/SCI/013). A.~V. is funded by FONDECYT post-doctoral Grant No.~3200884. The Centro de Estudios Cient\'{\i}ficos (CECs) is funded by the Chilean Government through the Centers of Excellence Base Financing Program of ANID.


\bibliographystyle{utphys}
\bibliography{Nuclear_biblio}

\end{document}